\documentclass[useAMS,hyperref]{mn2e}
\usepackage[dvips]{graphicx}
\usepackage{subfigure}
\usepackage{color}
\usepackage{amsmath}
\usepackage{amssymb}
\usepackage{textcomp}
\usepackage{natbib}

\newcommand{\msol}{M$_{\odot}$}
\newcommand{\lsol}{L$_{\odot}$}

\title[The G305 complex: a Hi-GAL Far-IR study]{The G305 star-forming complex: Embedded Massive Star Formation Discovered by Herschel Hi-GAL\thanks{{\it Herschel} is an ESA space observatory with science instruments provided by European-led Principal Investigator consortia and with important participation from NASA.}}
\author[A. Faimali et al.]
{A. Faimali\,$^{1}$\thanks{E-mail: a.faimali@herts.ac.uk},
M. A. Thompson\,$^{1}$, L. Hindson\,$^{1,2}$,  J. S. Urquhart\,$^{3}$, M. Pestalozzi\,$^{4}$, 
\newauthor 
S. Carey\,$^{5}$, S. Shenoy\,$^{6}$, M. Veneziani\,$^{5}$, S. Molinari\,$^{4}$, and J. S. Clark\,$^{7}$
\newauthor
\\$^{1}$Centre for Astrophysics Research, Science and Technology Research Institute, University of Hertfordshire, AL10 9AB, UK
\\$^{2}$ATNF, CSIRO Astronomy and Space Science, P.O. Box 76, Epping, NSW 1710, Australia
\\$^{3}$Max Planck Institut f\"{u}r Radioastronomie, Auf dem H\"{u}gel 69, 53121 Bonn, Germany
\\$^{4}$Instituto di Fisica dello Spazio Interplanetario, CNR, via Fosso del Cavaliere, I-00133 Roma, Italy
\\$^{5}$Spitzer Science Center, California Institute of Technology, M/S 220-6, 1200 E. California Blvd., Pasadena, CA 91125, USA
\\$^{6}$Space Science Division, NASA Ames Research Center, M/S 245-6, Moffett Field, CA 94035, USA
\\$^{7}$Department of Physics and Astronomy, The Open University, Walton Hall, Milton Keynes, MK7 6AA, UK}

\date{Released 2011 Xxxxx XX}

\pagerange{\pageref{firstpage}--\pageref{lastpage}} \pubyear{2011}

\def\LaTeX{L\kern-.36em\raise.3ex\hbox{a}\kern-.15em
    T\kern-.1667em\lower.7ex\hbox{E}\kern-.125emX}

\begin{document}

\label{firstpage}

\maketitle

\begin{abstract}

We present a \textit{Herschel} far-infrared study towards the rich massive star-forming complex G305, utilising PACS 70, 160\,\micron\,\,and SPIRE 250, 350, and 500\,\micron\,\,observations from the Hi-GAL survey of the Galactic plane. The focus of this study is to identify the embedded massive star-forming population within G305, by combining far-infrared data with radio continuum, H$_{2}$O maser, methanol maser, MIPS, and Red MSX Source survey data available from previous studies. By applying a frequentist technique we are able to identify a sample of the most likely associations within our multi-wavelength dataset, that can then be identified from the derived properties obtained from fitted spectral energy distributions (SEDs). By SED modelling using both a simple modified blackbody and fitting to a comprehensive grid of model SEDs, some 16 candidate associations are identified as embedded massive star-forming regions. We derive a two-selection colour criterion from this sample of log\,(F$_{70}$/F$_{500}$)\,$\geq\,1$ and log\,(F$_{160}$/F$_{350}$)\,$\geq\,1.6$ to identify an additional 31 embedded massive star candidates with no associated star-formation tracers. Using this result we can build a picture of the present day star-formation of the complex, and by extrapolating an initial mass function, suggest a current population of $\approx\,2\,\times\,10^{4}$ young stellar objects (YSOs) present, corresponding to a star formation rate (SFR) of 0.01\,-\,0.02 M$_{\odot}$\,\,yr$^{-1}$. Comparing this resolved star formation rate, to extragalactic star formation rate tracers (based on the Kennicutt-Schmidt relation), we find the star formation activity is underestimated by a factor of $\geq$\,2 in comparison to the SFR derived from the YSO population.

\end{abstract}

\begin{keywords}
stars: formation - infrared: ISM - HII regions - methods: statistical
\end{keywords}

\section{Introduction}

\begin{figure*}
\vspace{-14pt}
\begin{center}
\includegraphics[width=1.1\textwidth]{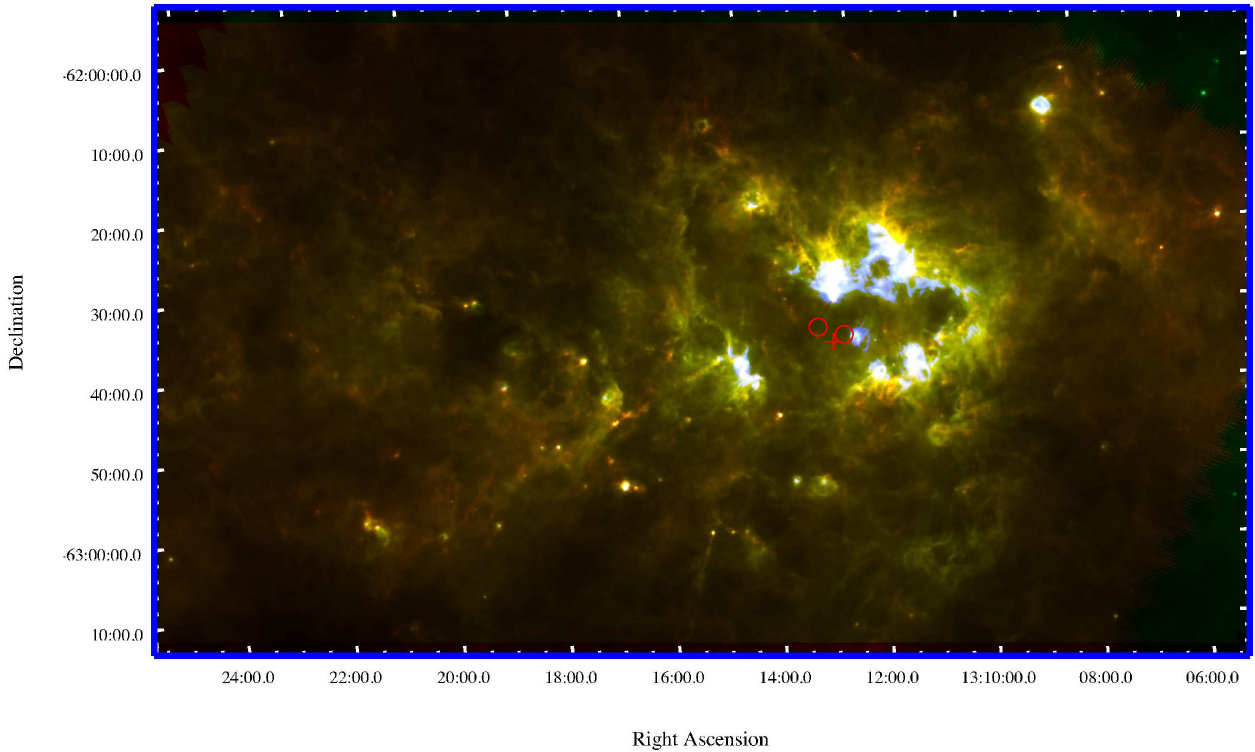} 
\vspace{-15pt}
\caption{Three-colour (blue\,=\,70\,$\mu$m, green\,=\,160\,$\mu$m, red\,=\,350\,$\mu$m) \textit{Herschel} Hi-GAL image of the giant HII region G305. The 70\,\micron\,\,emission corresponds to hot dust emission ($\geq\,40$\,K), while the 350\,\micron\,\,emission emanates from more cold dust ($\leq\,20$\,K). The positions of both Danks 1 \& 2 (circles), and WR 48a (cross) are overplotted.}
\vspace{-10pt}
\label{Figure:G305}
\end{center}
\end{figure*}

The G305 star-forming complex is one of the most massive and luminous star-forming regions in the Galaxy \citep{Clark:2004vd, Hindson:2010jt}, centred on the two optically visible open clusters Danks 1 \& 2 \citep{Danks:1984xy}, and the Wolf-Rayet star WR 48a. Located inside the Scutum-Crux arm within the Galactic plane at $l$\,=\,305$^{\circ}$, $b$\,=\,0$^{\circ}$, and found at a distance of $\sim$\,4\,kpc, it has a projected diameter of $\sim$\,30\,pc, and an estimated age of some 3\,-\,5 Myr \citep{Clark:2004vd}. Coincident with G305 are numerous signposts of massive star formation, some of which are located on the periphery of the central cavity, such as infrared hotspots, compact and ultra-compact (UC) HII regions, H$_{2}$O, OH, and methanol masers \citep{Urquhart:2007sc, Hindson:2010jt}.

We define a massive star as a star with sufficient mass to produce detectable radio emission associated with a HII region (i.e. M\,$>$\,8\,M$_{\odot}$). They play a key role in the Universe; their presence has a profound effect on the stellar and planetary formation process, while also on the physical, chemical, and morphological structure of galaxies \citep{Kennicutt:2005ni}. Their UV radiation output ionises the surrounding interstellar neighbourhood providing the principle source of heating in the interstellar medium (ISM), while also having an impact on subsequent star formation. Through various process, such as stellar winds and supernovae, their mechanical energy output serves as the energy and momentum inputs to the surrounding ISM, helping to sculpt the structure and energetics of the ISM, and hence in host galaxies \citep{Zinnecker:2007mk}. 

Though they are important, in the shaping and evolution of their host galaxies, the physics of the formation and evolution of massive stars is unclear. Firstly, there is the issue of high dust extinction making direct observation difficult; sites of massive star formation tend to be embedded within very opaque cloud cores, which suffer from visual extinction greater than 100 magnitudes \citep{Bally:2005zf}. Related to this also, is the fact that massive stars are predominantly found to form within dense, stellar clusters \citep{de-Wit:2005fk}. In all, this makes for a particularly difficult challenge in being able to study the origins of massive stars.

Fig.\ref{Figure:G305} underlines the dynamic morphology of the region, with the central location  Danks 1 \& 2 and WR 48a thought to be the main source of energy input and the driving force behind the expansion and clearing of the central diffuse HII region in the complex \citep{Clark:2004vd}. The suggestion is an interaction between the centrally embedded sources and the surrounding cloud, with an occurrence of ongoing massive star formation being located in the hot dust emission sites (seen in blue in Fig.\ref{Figure:G305}) on the periphery of the central cavity. With the presence of numerous and different epochs of star formation in one location, and the relative close proximity, the G305 star-forming complex affords us an exceptional opportunity to study the nature of massive star formation, and investigate the environmental impact this may have on the formation of future generations of stars \citep{Elmegreen:2002dg, Elmegreen:1977bs}.

This paper follows in a series of previous studies of massive star formation within the G305 star-forming complex. Previously, \cite{Hindson:2010jt} have focused on the reservoir for star formation within the region through observations of NH$_{3}$ emission, while tracing sites of active star formation through H$_{2}$O maser emission. To complement this, \cite{Davies:2011gl} have recently carried out a near-infrared study of the two central open clusters Danks 1 and 2; while current work has focused on identifying compact radio emission that is indicative of UC HII regions towards G305 \citep{Hindson:2012lr}. 

In this paper, we present a far-infrared (far-IR) study of the G305 complex using \textit{Herschel} \citep{Pilbratt:2010eg}, in conjunction with radio continuum, H$_{2}$O maser, methanol maser, MIPS, and Red MSX Source survey data, with the aim of identifying sites of embedded massive star formation. In this study we are able to resolve the embedded population within G305, since the \textit{Herschel} far-IR observations are unaffected by dust, and combined with mid-IR data, constrain the luminosities of individual YSOs. By incorporating the luminosities of the embedded massive star-forming population with the initial mass function (IMF), we are able to determine the SFR of G305 and investigate the star formation history of the region. This resolved, Galactic SFR can then be compared to extragalactic SFR indicators to test whether the two regimes are consistent with one another, and identify where fundamental differences may lie \citep{Heiderman:2010io, Lada:2010tb}. 

The present study serves as an example of how \textit{Herschel} data can be applied to Galactic star-forming regions, such as G305, in order to identify the high-mass stellar content of such complexes, and how the star formation activity can be inferred from this population. Following from this, a comprehensive YSO counting approach, similar to that conducted by \cite{Povich:2011oj} for the Carina complex, will be conducted to tackle the incompletenesses present at the intermediate, and low-mass range in a following paper by the authors. By combining \textit{Spitzer} GLIMPSE \citep{Benjamin:2003qy}, and VISTA VVV \citep{Minniti:2010fk} data alongside our present datasets we aim to conduct a complete census of the YSO population of G305, studying the physical parameters of these sources, their evolutionary stage, and spatial distribution within the complex.

This paper is organised into the following sections; in Section 2 we present and discuss the details of the observations and the data analysis, in Section 3 we present methods to identify sources detected in the far-IR through consideration of their spectral energy distribution, in Section 4 we present our discussion and place G305 in context with other star-forming regions, and in Section 5 we summarise our results.

\section{Observations \& Data Analysis}
\subsection{Herschel Hi-GAL}

The \textit{Herschel} Infrared GALactic plane survey (Hi-GAL) \citep{Molinari:2010gl} is an Open Time Key Project on board the ESA \textit{Herschel} Space Observatory \citep{Pilbratt:2010eg},  mapping a two degree wide strip of the inner Galactic plane, in the longitudinal range $\vert\,l\,\vert\leq\,60^{\circ}$ and latitude range $\vert\,b\,\vert\leq\,1^{\circ}$. The survey takes advantage of the PACS \citep{Poglitsch:2010mw} and SPIRE \citep{Griffin:2010jw} cameras operating in parallel mode, scanning the sky in a raster fashion at a rate of 60\,$\arcsec$\,s$^{-1}$, while PACS and SPIRE acquire data simultaneously. The survey has a total angular resolution, $\theta$, of $10\arcsec\,\leq\,\theta\,\leq\,30\arcsec$\, across five photometric bands at 70 and 160\,$\mu$m with PACS and 250, 350, and 500\,$\mu$m with SPIRE. The overall aim is to catalogue star-forming regions and study cold structures across the ISM. Using the broad spectral coverage available, the intention is to study the early phases of star formation, with particular focus on providing an evolutionary sequence for the formation of massive stars within the Galactic plane.

The catalogue of compact infrared sources for the G305 region is obtained using the CuTEx (CUrvature Thresholding EXtractor) code highlighted by \cite{Molinari:2011ws}. The detection technique considers the curvature properties of astronomical images, rather than source detection through signal intensity, by building a ``curvature" image from the observed image using double-differentiation in four separate directions. The advantage of this approach is that resolved, compact sources are easily detected, while the diffuse thermal emission from cold dust associated with the fore/background is greatly reduced. Photometry estimates of candidate sources are then performed by fitting an elliptical two-dimensional Gaussian with an underlying planar inclined plateau. 

The completeness of the infrared source catalogue is estimated using a simulated field of some 1000 artificial sources, initially at 250\,\micron, comprising both a compact dust component (such as YSO envelopes, or dense molecular cores or clumps), and the diffuse emission present towards the Galactic plane \citep{Molinari:2010my}. In all, source recovery rate is found to be 90\% for sources of peak fluxes of 0.2\,Jy/pixel (equivalent to approximately 5$\sigma$), with peak fluxes being within 30\% of their input value in 80\% of recovered input sources \citep{Molinari:2011ws}.

Identifying each source within the five different bands is obtained considering their basic positional association, starting with source extraction at the longest of the observed wavelengths, i.e. at 500\,\micron. From this first band, an association is established with the next band (i.e. 350\,\micron) if a source is present within a search radius that corresponds to the \textit{Herschel} beam size at the longer of the two considered wavelength, in this case a radius of 18.1\arcsec\ at 500\,\micron. In the case where multiple associations are found, the closest one is kept. This is more frequent at 70\,\micron\,\,\citep{Elia:2010yp}. At this stage the flux contribution at longer wavelengths has not been split up into contributions from multiple associations at 70\,\micron, however we find 87\% of 500\,\micron\ sources are associated with a single 70\,\micron\ detection. Of the remaining 500\,\micron\ sources with multiple 70\,\micron\ associations, the majority are found to have the bulk of the 70\,\micron\ flux assigned to their primary 70\,\micron\ counterpart. We find only 3\% of 500\,\micron\ sources to have lost a fraction of 70\,\micron\ flux through multiple associations. This loss is deemed negligible, being on the order of 1.5\% of the total 70\,\micron\ flux.

Finally, the celestial coordinates assigned to the sources are those which correspond to the shortest wavelength association, which by definition will have the highest spatial resolution \citep{Molinari:2010my}. As a result of this, in total some 3288 infrared sources are detected across the G305 region in Hi-GAL; of these some 1913 sources are detected at 70\,\micron, 1658 at 160\,\micron, 1257 at 250\,\micron, 856 at 350\,\micron, and 530 sources at 500\,\micron. Not all sources are detected in all wavebands, and this can be accounted for by either positional association failing, or that the flux obtained for a source in a particular band had been corrupted by factors such as source crowding, or confused background conditions \citep{Molinari:2010my}.

\begin{figure}
\begin{center}
\includegraphics[width=0.4\textwidth,angle=-90]{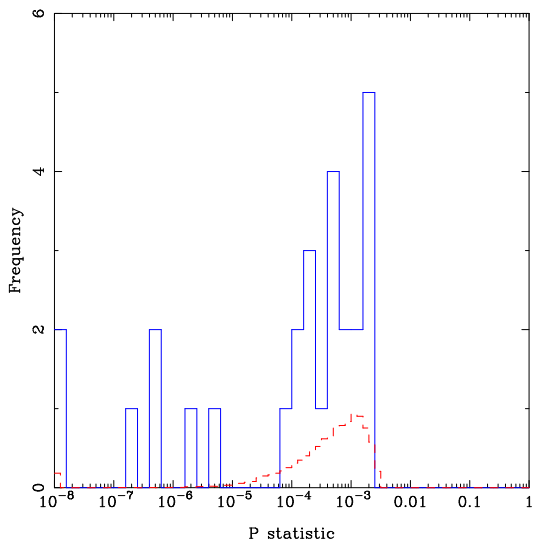} 
\caption{Distribution of P-statistics of identified associations (solid blue line), against the distribution of P\,\arcmin-statistics obtained from Monte Carlo associations (dashed red line).}
\label{Figure:P-Statistic}
\end{center}
\end{figure}

\subsection{Ancillary Data}

To complement our data set we utilise mid-IR, radio, H$_{2}$O and methanol maser data matched to the Hi-GAL far-IR data to broaden our view of the G305 complex, aiding in the consideration of the morphology, and star formation within the region.

We firstly make use of the 5.5\,GHz radio continuum observations towards G305 using the Australia Telescope Compact Array (ATCA) \citep{Hindson:2012lr} as a means to distinguish UC HII regions across the complex. A key method of detecting HII regions is through the associated radio emission; at wavelengths greater than 3\,mm, thermal dust emission drops off, and the dominant detection is due to thermal free-free radio emission from the ionised gas in the inner region surrounding the massive star. The Lyman continuum photons that are emitted from the central star are balanced by recombination within the volume of ionised hydrogen, producing the detected radio emission. For this emission, stellar models suggest stars of spectral type B3 or earlier are capable of producing the required ionising flux of Lyman continuum photons (E\,$>$\,13.6\,eV, i.e. 912\,$\AA$) \citep{Crowther:2003qe}. Such a spectral type is selected since the Lyman continuum flux then begins to rapidly drop off with decreasing effective temperatures \citep{Panagia:1973dm}.

To also locate areas of ongoing star formation, we employ both 22\,GHz H$_{2}$O maser observations using the Australia Telescope National Facility (ATNF) telescope Mopra \citep{Hindson:2010jt}, and the 6.7\,GHz methanol maser observations of the Methanol Multi-Beam survey \citep{Green:2009jv, Caswell:2010rb} with the Parkes 64\,m radio telescope and follow ups by the ATCA. Since their discovery, methanol masers have been recognised as one of the most distinct signposts of massive star formation \citep{Menten:1991tg}, and only found close to high-mass young stars \citep{Minier:2003xf}. H$_{2}$O maser emission, on the other hand, has also been shown to provide a useful indicator of both low- and high-mass star formation, and are found to associate mainly with both hot molecular cores and UC HII regions \citep{Furuya:2003xf}.

Finally, we also make use of mid-IR data from both the Red MSX Source (RMS) survey that aims to identify a large sample of genuine massive young stellar objects (MYSO) and UC HII regions located throughout the Galactic plane \citep{Urquhart:2008xr}, and the 24\,\micron\ point source catalogue from the \textit{Spitzer} MIPS Galactic Plane Survey (MIPSGAL) \citep{Mizuno:2008fk, Carey:2009qa} which aims to identify all high-mass protostars located in the inner Galactic disk.

\subsection{Identification Procedure}

Our aim is to identify a statistically reliable sample of counterparts to the Hi-GAL infrared catalogue, by matching this catalogue to the ATCA radio, MMB, RMS, Mopra H$_{2}$O maser, and MIPSGAL catalogues. By using a frequentist technique, highlighted by \cite{Lilly:1999qd}, we can identify all associations found within G305, by establishing the probability that matches to Hi-GAL sources are not the result of a chance alignment. We approach this using a Monte Carlo method, with the technique outlined in the following subsection for the case of Hi-GAL infrared and ATCA radio counterparts.

\subsubsection{Association Probabilities}

Starting from the matching of both the Hi-GAL infrared and ATCA radio catalogues, we need to consider the possibility of chance alignments and provide an estimate as to how reliable each individual match may or may not be. We therefore need to derive some statistical argument that considers the probability that a candidate compact radio source is indeed correctly identified within the search radius of the associated infrared source. A method based on the positional coincidence, similar to that adopted by \cite{Downes:1986gn} and \cite{Sutherland:1992iw} is employed.

The necessity to work out a statistic which can inform whether an association is indeed true, or by chance can be shown by considering the ATCA radio data. A certain number of the 71 radio sources identified across the field will be accounted for by extragalactic background sources. We can estimate the total amount of background sources empirically from the extragalactic source count approach of \cite{Anglada:1998tr}. The number of background radio sources, $\langle\,N_{Radio}\,\rangle$, that would be expected to be observed given the region size, and frequency of observation is defined by:

\begin{equation}
\langle\,N_{Radio}\,\rangle\simeq\left(\frac{\theta_{f}}{\theta_{A}}\right)\,1.1\,S_{0}^{-0.75}
\label{eq:Background Radio}
\end{equation}
where $\theta_{f}$ is the diameter of our observed field, $\theta_{A}$ is the FWHM of the primary beam (in arcminutes), and $S_{0}$ is the sensitivity of the radio observations at 5.5\,GHz. Using this approach, we find that some 60\,$\pm$\,8 background sources in total should be detected across the G305 field \citep{Hindson:2012lr}.

As an example, to explain the positional coincidence method, we use our identified infrared-radio associations. For each association we calculate that for each ATCA compact radio source candidate with magnitude \textit{m} at a distance \textit{r} from the matched to the Hi-GAL infrared source, there is a surface density \textit{N$_{\sigma}$} of radio sources brighter than \textit{m} across the G305 field. We therefore obtain the mean number \textit{$\mu$} of chance sources that are closer and brighter than the matched to candidate ATCA radio source:

\begin{equation}
\mu=\pi r^{2}N_{\sigma}
\label{eq:P-stat}
\end{equation}
We utilise this statistic to suggest the fraction of sources within a sample size of \textit{n}, that we would expect to have an incorrect candidate association identified within a matching radius \textit{r}. In a sense, the probability of the association is:

\begin{equation}
P=[1-\exp(-\mu)]\approx\mu\,\,\rm{for}\,\, \mu\ll1
\label{eq:P-statistic}
\end{equation}
From this, we find that if the P-statistic, \textit{P}\,$\ll$1 for an individual source, then that particular identified
association is unlikely to be the result of a chance association.

However this statistic alone does not provide us with a firm argument to the reliability of associations; merely the chance that the particular source would have an association within the specific matching radius. Rather, a more reliable means to measure whether an individual association can be deemed correct, is to compare the number of identifications in the total sample with a particular P-statistic, against the amount of associations, \textit{nP}, that would have been expected from a randomised association between the infrared and radio populations.

To do this, we run 5000 Monte Carlo simulations that follow the previous matching criteria between both Hi-GAL and ATCA catalogues; where the positions of each compact radio source have been randomized, with the only constraint being that they remain within the field. From these Monte Carlo results, we can then calculate the P\,\arcmin-statistic for each of the identified Monte Carlo candidate matches. By comparing the P-statistic of our associations to those P\,\arcmin-statistics from the randomised sample, we examine the ratio of sources with a particular value of \textit{P} to the similar value of \textit{P\,\arcmin} from the spurious identifications. If this ratio is high, i.e. there are many associations with a particular value of \textit{P} compared to that of \textit{P\,\arcmin}, then we can mark that particular association as a secure identification. This is shown for a matching radius of 30\arcsec\ in Fig. \ref{Figure:P-Statistic}.

\begin{figure}
\begin{center}
\includegraphics[width=0.4\textwidth,angle=-90]{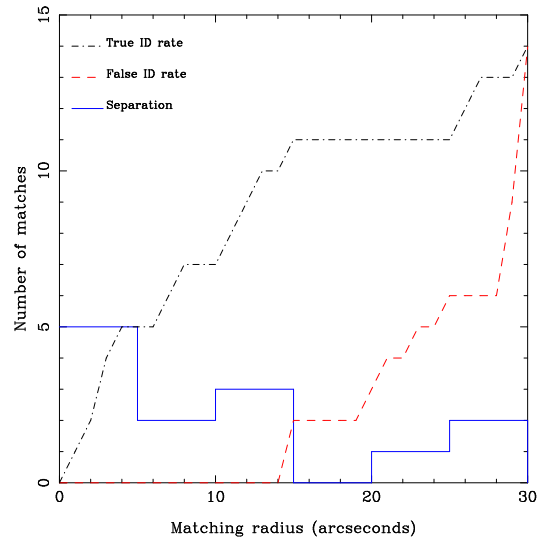} 
\caption{Distribution of radial offsets between Hi-GAL sources and ATCA radio counterparts (open histogram). The dashed line gives the expected cumulative number of false radio IDs as a function of the separation cut-off, while the dot-dashed line shows the cumulative number of true radio IDs for each separation cut-off. In this context, the optimum matching radius is found to be 15\arcsec.}
\label{Figure:Optimum Cut-off}
\end{center}
\end{figure}

\subsubsection{Separation Cut-off}

\begin{table}
\scriptsize
\addtolength{\tabcolsep}{-4pt}
\caption{Optimum matching radius for each data set to the Hi-GAL G305 field.}
\begin{centering}
\begin{tabular}{lccccc}
\hline
Data Set & True ID & False ID & Optimum Matching Radius (\arcsec)\\
\hline
ATCA & 11 & 2 & 15\\
MMB & 14 & 2 & 15\\
RMS & 9 & 2 & 5\\
H$_{2}$O Masers & 4 & 1 & 15\\
MIPS & 500 & 63 & 10\\
\hline
\end{tabular}
\label{tab:Catalogue Matching Radius}
\end{centering}
\end{table}

A useful outcome from the use of P-statistics is that it allows for the determination of the separation cut-off for each dataset; essentially the maximum matching radius at which one considers the majority of associations found to be reliable. The selection of the separation cut-off depends on several factors. Clearly a smaller cut-off gives rise to fewer counterparts, increasing the likelihood of missing a true counterpart. Conversely, a larger cut-off increases the risk of associating to a very bright but unrelated counterpart. From the P-statistics we are then able to derive the optimum matching radius for the data using the technique highlighted by \citet{Dye:2009ib}. Firstly, we match between the Hi-GAL and ATCA data at a minimum matching radius of 1\arcsec. By then comparing the P-statistic of our associations to those P\arcmin-statistics from the randomised sample, we examine the ratio of sources with a particular value of \textit{P} to the similar value of \textit{P\,\arcmin} from the spurious identifications. If this ratio is high, i.e. there are many associations with a particular value of \textit{P} compared to that of \textit{P\,\arcmin}, then we can suggest that particular association as a secure identification. This process is stepped through, at intervals of 1\arcsec\, up to a maximum matching radius of 30\arcsec\ . As a result of this we are then able to compute a true and false ID rate as a function of separation.

The result is shown in Fig. \ref{Figure:Optimum Cut-off}, with the true ID rate, the false ID rate, and the separation between Hi-GAL and ATCA true associations in the G305 field shown as a function of the separation cut-off. From this we can see that the true ID rate levels out at 15\arcsec\ (with a total of eleven true associations), and only increases again at 25\arcsec\ onward (with a further three true associations being found).  These additional three true associations found at radii $\geq$\,25\arcsec\ correspond to Hi-GAL matches to the brightest ATCA radio sources found in the G305 field. In choosing the search radius, we wish to maximize the number of secure, unambiguous identifications and to minimize the number of real counterparts missed. It is not seemingly clear as to why there is a well-defined P-statistic peak at a lower matching radii; being due to faint, real radio identifications, contesting (in terms of low P-statistic) with rare, brighter sources, which are unrelated to the Hi-GAL source (such as background radio galaxies). By increasing our matching radius, any correct identifications will tend to drop off, with the levelling out in true identifications being balanced by those bright unrelated radio sources present; since they are very bright, their P-statistic tends to zero as soon as they are within the matching radius. Therefore allowing a matching radius well beyond 15\arcsec\ is not suggested, as the amount of secure associations tend to level out, while matches to contaminant radio sources will preserve the number of perceived ‘true’ associations \citep{Ivison:2010mh}.

We also note from Fig. \ref{Figure:Optimum Cut-off} that the false ID rate only becomes noticeable for 15\arcsec\ onwards, with two possible false IDs found at 15\arcsec. Finally, the separation between Hi-GAL and ATCA true associations, shown in the open histogram, emphasises the fact that the majority of true associations are found within a matching radius of $\leq$\,15\arcsec. The outcome of the matching procedure returns an optimal separation cut-off of 15\arcsec\ for the radio counterparts in the Hi-GAL field. 

Suggesting that a radio source some 15\arcsec\ from an IR source is a secure match may seem counter-intuitive, but two factors need to be taken into consideration; the \textit{Herschel} beam size, and the physical nature of the IR sources. Our goal is the identification of a sample of embedded massive star-forming regions that themselves are extended in nature, on average around 29\arcsec\ in size, and with the beam size some 18\arcsec\ at 250\,\micron\ \citep{Traficante:2011bo}, a secure match within 15\arcsec\ is acceptable. 

The approach is repeated for the MMB, RMS, H$_{2}$O Masers, and MIPS data sets, with the total number of associations found and the optimum matching radius shown in Table \ref{tab:Catalogue Matching Radius}. The result of the technique is a statistically robust sample of true associations that have been identified over multiple datasets, which can later be investigated and analysed.

\section{Spectral Energy Distributions of Sources}

To obtain estimates of the physical properties of our associations, the observed SEDs were fitted with firstly a simple modified blackbody, incorporating the \textit{Herschel} Hi-GAL observations at 70, 160, 250, 350, and 500\,\micron. We then can obtain estimates of the bolometric luminosity of each association, by broadening the SED coverage with both the MSX 21 and MIPS 24\,\micron\ fluxes, and fitting to the grid of SED models from \cite{Robitaille:2006mo}. We find that the modified blackbody fails to reproduce the observed flux at $\lambda$\,$\leq$\,70\,\micron, hence associations with solely \textit{Herschel} detections are best suited to modified blackbody fits. Those associations with detections at 21 and 24\,\micron, suggesting a warm embedded YSO, are more reliably reproduced with the  \cite{Robitaille:2006mo} SED models, that incorporate both the central embedded source, and the surrounding envelope.

\subsection{Modified Blackbody Fitting}

\begin{figure}
\begin{center}
\includegraphics[width=0.4\textwidth,angle=-90]{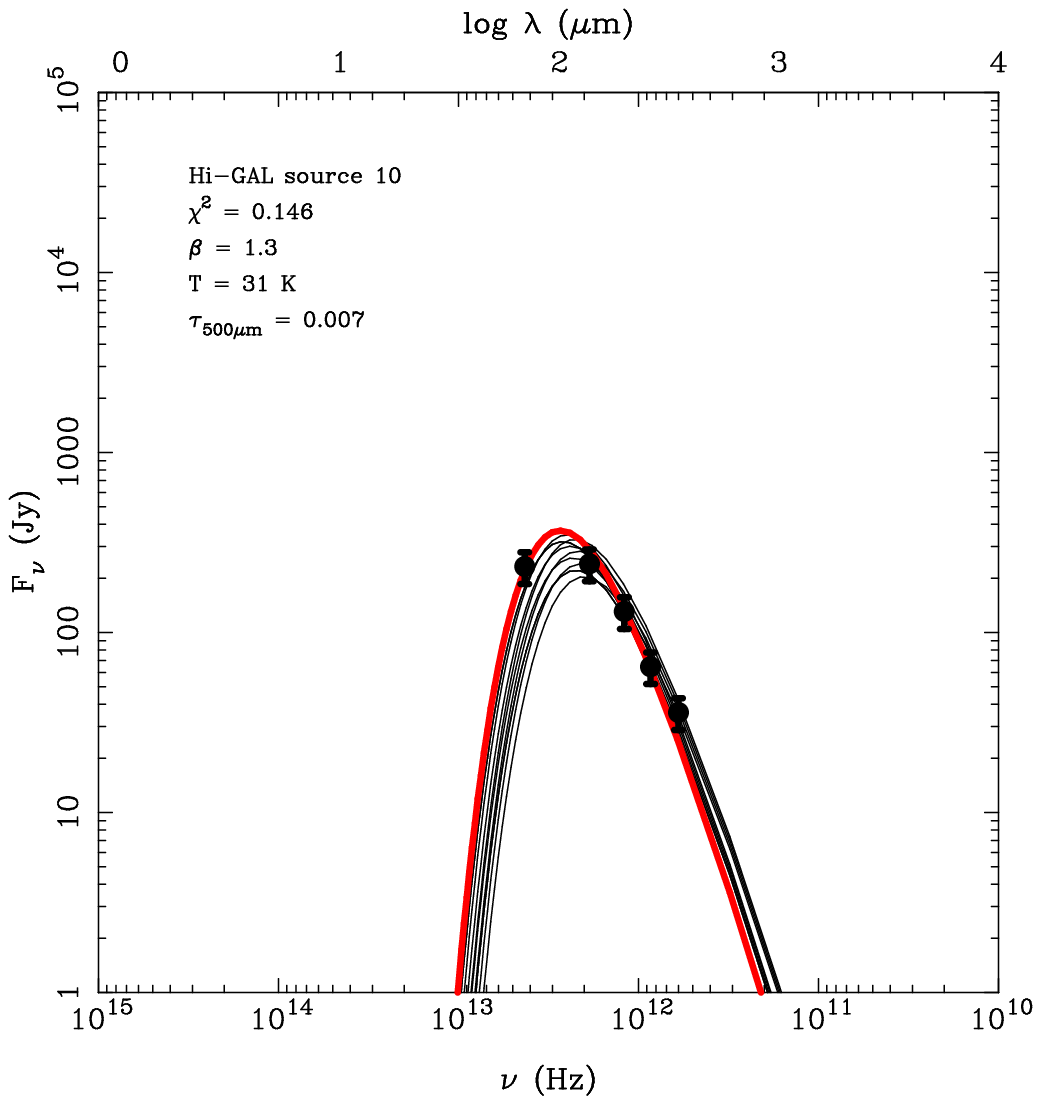}
\includegraphics[width=0.4\textwidth,angle=-90]{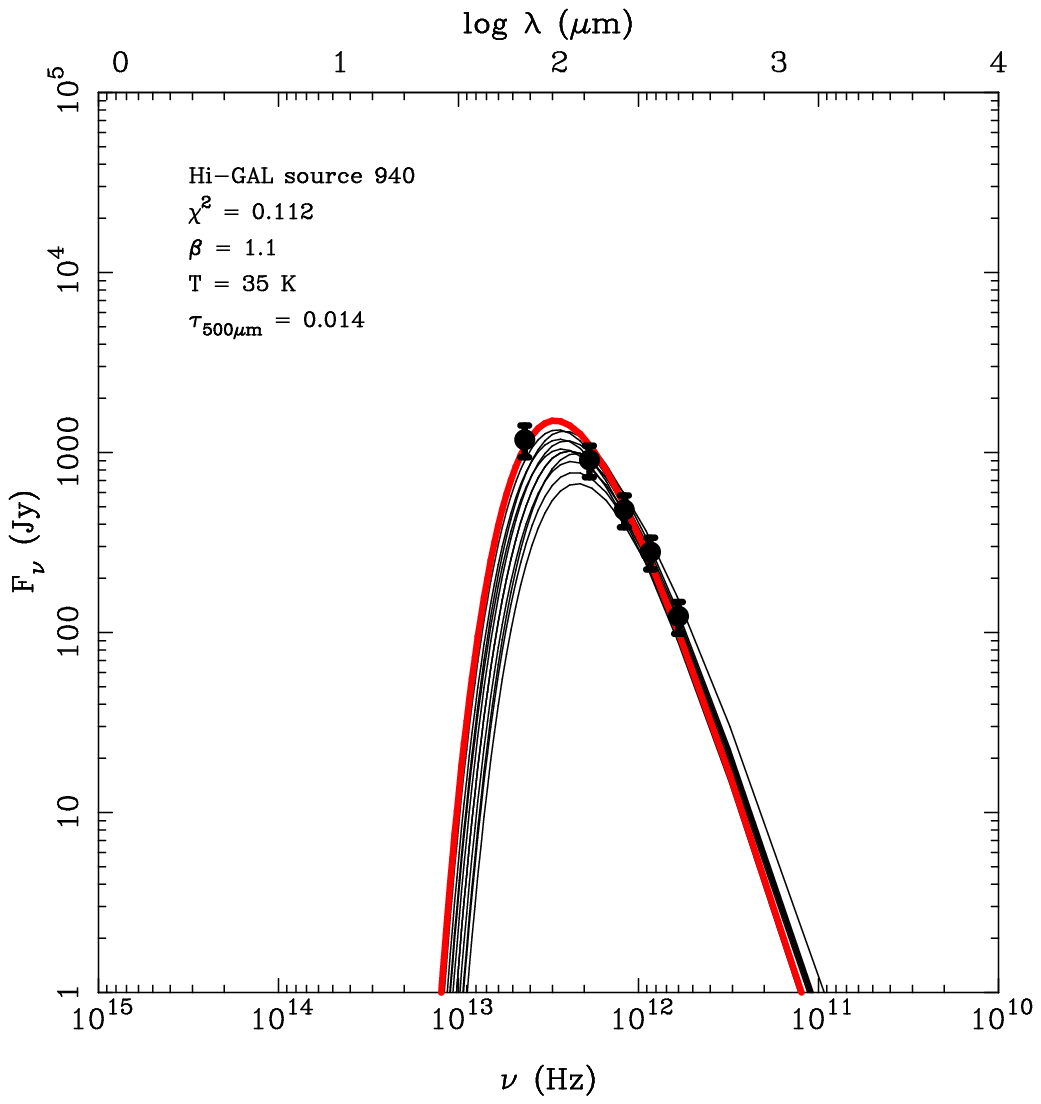}
\end{center}
\caption[dum]{Fits to two associations using a modified blackbody SED. Physical properties derived from the fit are shown, along with minimum $\chi^{2}$ for the best fit. The red solid line represents the best fit model, while the grey solid lines represent other models with a good fit to the data.}
\label{Figure:Greybody SED Fits}
\end{figure}

In order to firstly be able to derive the basic physical properties for each of our associations such as dust temperature and emissivity index, it was necessary to model the SEDs with a modified blackbody. We are justified in taking the approach of a simple modified blackbody, since the wavelength coverage measured by both PACS and SPIRE trace the peak of the dust SED. Far-IR emission is due to large dust grains (15\,-\,100\,$\mu$m), which are more stable and tend to dominate the total dust mass, while also tending to trace all phases of the ISM. Therefore, measurements in Hi-GAL will be most sensitive to temperature variations, while also providing an accurate tracer of the overall ISM column density. 

Since we have a embedded massive star within a dust cocoon we cannot assume a blackbody, but rather a modified blackbody that takes into consideration the optical depth, and the dust emissivity, as clearly emission is not optically thick at all frequencies. Thus we adopt a modified blackbody fit to the SED of the form \citep{Ward-Thompson:1990et}:

\begin{equation}
F_{\nu}=\Omega\,B_{\nu}(T)(1-e^{-\tau})\,\,\rm{[Jy]}
\label{eq:Modified Blackbody}
\end{equation}
where $\Omega$ is the effective solid angle of the source, in sr, B$_{\nu}$(T) is the Planck function, T is the dust temperature, and $\tau$ is the optical depth such that:

\begin{equation}
\tau=\left(\frac{\nu}{\nu_{c}}\right)^{\beta}
\label{eq:Opacity}
\end{equation}
where $\beta$ is the dust emissivity index, and $\nu_{c}$ is the critical frequency at which $\tau\,=\,1.$

The free parameters that are derived from the fit include the dust temperature, the dust emissivity index, and the critical frequency. We also note that throughout the calculations each \textit{Herschel} flux was assigned an uncertainty of 20\%, this being the calibration error.

Fitting was performed via a $\chi^{2}$ minimisation, by considering the observed flux at each of the five Hi-GAL wavebands available for every individual association. The $\chi^{2}$ minimisation was such that \citep{Hunter:2000gc}:   

\begin{equation}
\chi^{2}=\sum_{n}\left[1-\left(\frac{F_{\nu_{n},\rm{model}}}{F_{\nu_{n},\rm{observed}}}\right)\right]^{2}
\label{eq:Minimised Chi}
\end{equation}
where the ratio of the model flux to the observed flux is used to give equal weighting to each different wavelength regime.

With these parameters we can then derive the total (gas\,+\,dust) mass for each association using both the \textit{Herschel} fluxes at 500\,$\mu$m, and the corresponding free parameters derived from the minimum $\chi^{2}$ fit. We note that dust masses were calculated with an opacity $\tau_{500\mu m}$, since at this wavelength all sources were found to be optically thin, allowing for us to sample the dust at all depths. If dust masses were calculated at shorter wavelengths, where some sources were found to be optically thick, we would simply be tracing the dust distribution in the outer layers, and not in fact determining a total dust mass value. Therefore, by following the method highlighted in \cite{Hildebrand:1983yc}:

\begin{equation}
M=\frac{F_{\nu}D^{2}}{B_{\nu}(T)}C_{\nu}
\label{eq:Core Mass}
\end{equation}
where $D$ is the source distance, and $C_{\nu}$ is the mass coefficient (a factor that combines dust opacity and gas-to-dust ratio) \citep{Kerton:2001wj}:

\begin{equation}
C_{\nu}=\frac{M_{\rm{g}}}{M_{\rm{d}}\kappa_{\nu}}
\label{eq:Mass coefficient}
\end{equation}
where $M_{g}$ is the gas mass, $M_{d}$ is the dust mass, and $\kappa_{\nu}$ is the dust opacity. A value of the mass coefficient of C$_{\nu}=$ 50\,g\,cm$^{-2}$ at 850\,$\mu$m was initially chosen from those quoted in \cite{Kerton:2001wj}, assuming a gas to dust ratio of 100 and a dust emissivity index $\beta$ of 2. This value was then scaled to a value at 500\,$\mu$m, by adopting the value of $\beta$ from the minimum $\chi^{2}$ fits, and assuming that:

\begin{equation}
C_{\nu}\propto\nu^{-\beta}
\end{equation}

Fig. \ref{Figure:Greybody SED Fits} shows the result of the modified blackbody model SED fit for a sample of associations, with both the best fit model and those models deemed good fits from the criterion $\chi^{2}$\,/\,N$_{data}\,\leq$\,2 \citep{Povich:2011oj} shown. Included are the corresponding free parameters of the minimum $\chi^{2}$ fit, being the dust temperature, emissivity index,  and source opacity at 500\,$\mu$m ($\tau_{500\mu m}$).

\subsection{SED Fitting}

\begin{figure}
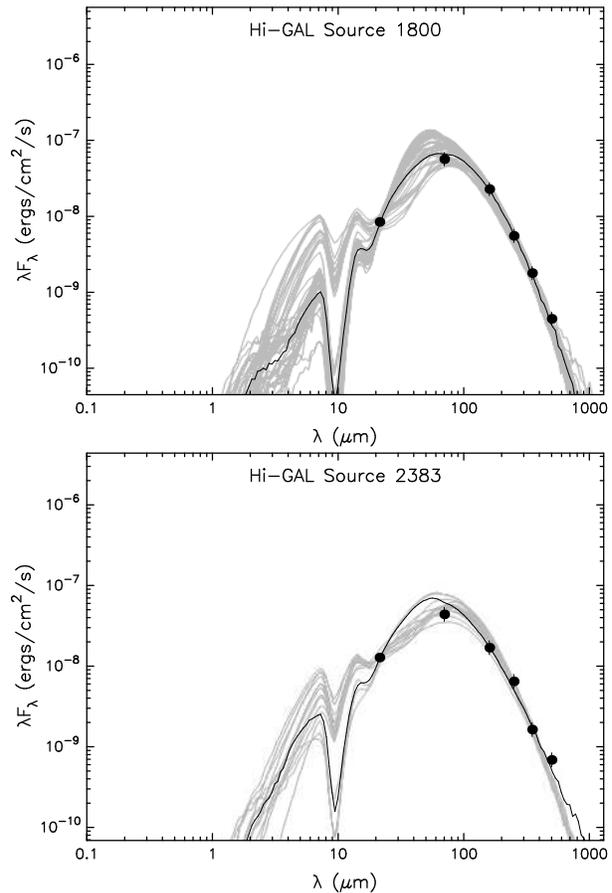

\begin{center}
\includegraphics[width=0.45\textwidth]{Source_1800.eps}
\includegraphics[width=0.45\textwidth]{Source_2383.eps}
\end{center}
\caption[dum]{Fits for two associations using the \cite{Robitaille:2007nv} fitting technique. Black solid line represents the best fit model, with the grey solid lines showing all other models providing a good fit to the data. The black filled circles show the Hi-GAL and MSX/MIPS data with error bars shown.}
\label{Figure:Robitaille SED Fits}
\end{figure}

We next obtain estimates of the bolometric luminosity by fitting the observed SEDs with the grid of young stellar object (YSO) model SEDs of \cite{Robitaille:2006mo}. Since the SEDs of embedded YSOs tend to peak at 100\,\micron\,,we require photometric data at $\lambda\,\geq\,10$\,\micron\,\,\citep{Mottram:2011ii, Povich:2011oj}, hence to broaden the SED coverage we incorporate the 21 and 24\,\micron\ fluxes obtained from MSX and MIPS. For these sources, the standard modified blackbody adopted earlier will fail to produce accurate SEDs at wavelengths $\lambda$\,$\leq$\,70\,\micron, thus a multi-component fit using the model SEDs of \cite{Robitaille:2006mo} that are then fit with the online SED fitting tool of \cite{Robitaille:2007nv}, that are based on the YSO/disk/envelope models of \cite{Whitney:2003zr} are employed.

The model grid of \cite{Robitaille:2006mo} consists of some 200,000 model SEDs incorporating a vast range of possible evolutionary stages, from an embedded protostellar phase to pre-main-sequence stars with low-mass circumstellar disks. Fitting of models to the data is done by varying the visual extinction, A$_{V}$, for a number of distances $d$ between $d_{min}$\,-\,$d_{max}$, in order to determine the optimum SED model and parameters set. A range of 0\,-\,20 is selected for the visual extinction \citep{Leistra:2005an}, while an averaged distance of 3.2\,-\,4.4\,kpc derived from both the kinematic distance to G305 and the spectroscopic distance of Danks 1 \& 2 is used \citep{Davies:2011gl}. The optimum SED model is determined through linear regression, with the result shown in Fig. \ref{Figure:Robitaille SED Fits}, where all models that fitted with a $\chi^{2}$ value satisfying $\chi^{2}$\,-\,$\chi^{2}_{best}$\,$\leq$\,3\,$\times$\,$n_{data}$ shown also.

The majority of sources, with no 21, 24, or 70\,\micron\ emission, have fitted SEDs that peak at wavelengths $\lambda$\,$\geq$\,160\,\micron\ and are found to have an averaged temperature of $\approx$\,14\,K.  \cite{Bontemps:2010fk} characterise YSOs within the Aquila rift complex from pre-stellar sources on the presence of either a 24 or 70\,\micron\ counterpart, with this emission originating from warm dust within the inner regions of the YSO envelope. Moreover, such emission is unlikely the result of external heating from the interstellar radiation field producing a detectable 70\,\micron\ counterpart \citep{Giannini:2012fk}, making sources with 21, 24, or 70\,\micron\ emission likely protostellar sources. Indeed, \cite{Dunham:2008fk} show that 70\,\micron\ emission is closely correlated to the overall luminosity of protostellar sources. Using this selection criteria, the suggestion is that these identified sources correspond to an earlier stage (possibly pre-stellar) that are poorly fitted by a protostar embedded in a dust envelope, and are more suited to a simple modified blackbody fit, yielding values of dust mass, temperature, and emissivity index. 

\section{Discussion}

In this section we discuss the global distribution of the embedded massive star-forming population within the G305 complex, and the general method of identification of this population; through the combination of the SED morphology and derived parameters. Using these identified sources, we are able to comment on the present-day star formation activity of G305, and place it into the context of Galactic star formation.

\subsection{General Properties Of Sources Within G305}

We firstly consider the global properties of G305 obtained from SED fitting, by both modified blackbody and \cite{Robitaille:2006mo} YSO models, for each association identified within the G305 region. For our fitting we consider sources with detections of N$_{data}$\,$\geq$\,3 as acceptable to fit an SED to, since we have three free parameters in our modified blackbody fitting, sources with limited detections will unlikely produce a reliable SED; this leaves us with some 503 sources. From these fits, we deem sources with $\chi^{2}$\,/\,N$_{data}\,\leq$\,2 \citep{Povich:2011oj}, as having a reliably fitted SED, yielding a total of 359 well fitted sources.

\begin{figure}
\mbox{\subfigure{\includegraphics[width=0.23\textwidth,angle=-90]{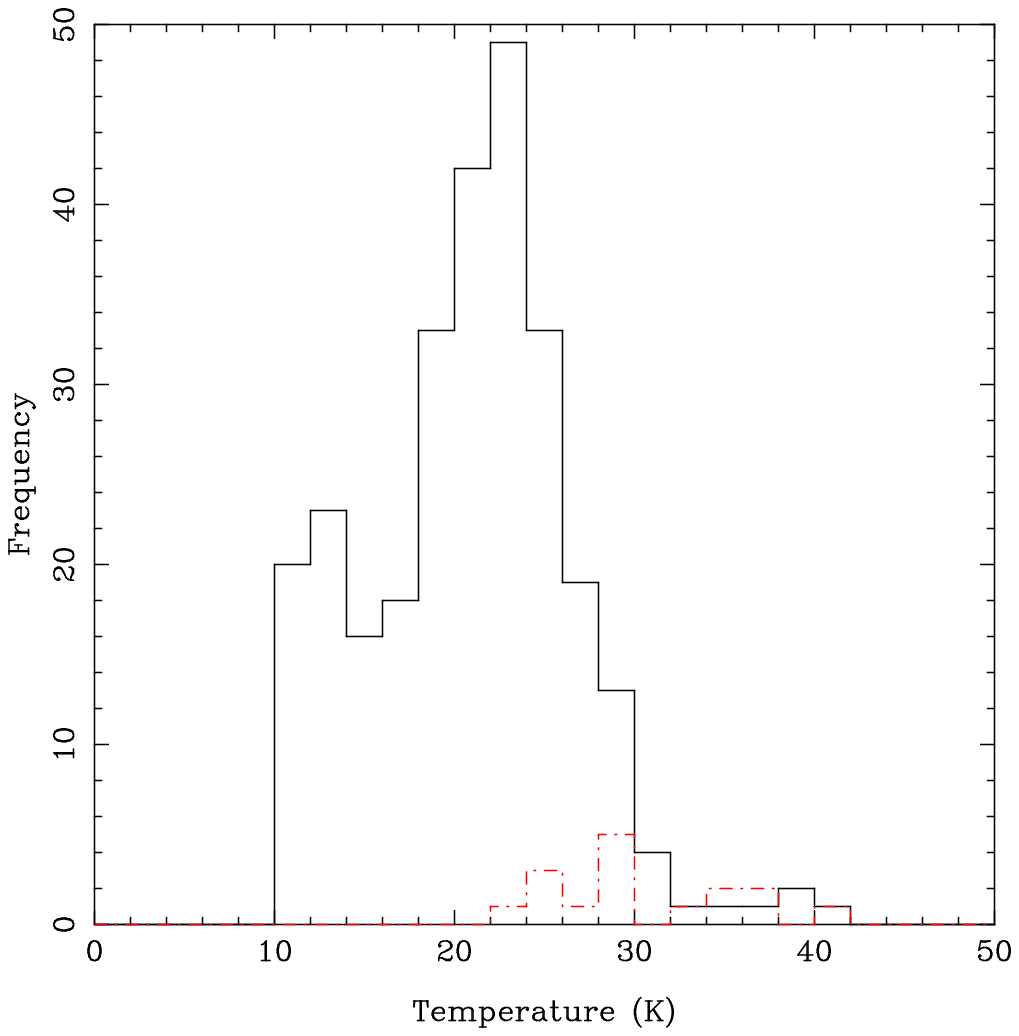}}
\subfigure{\includegraphics[width=0.23\textwidth,angle=-90]{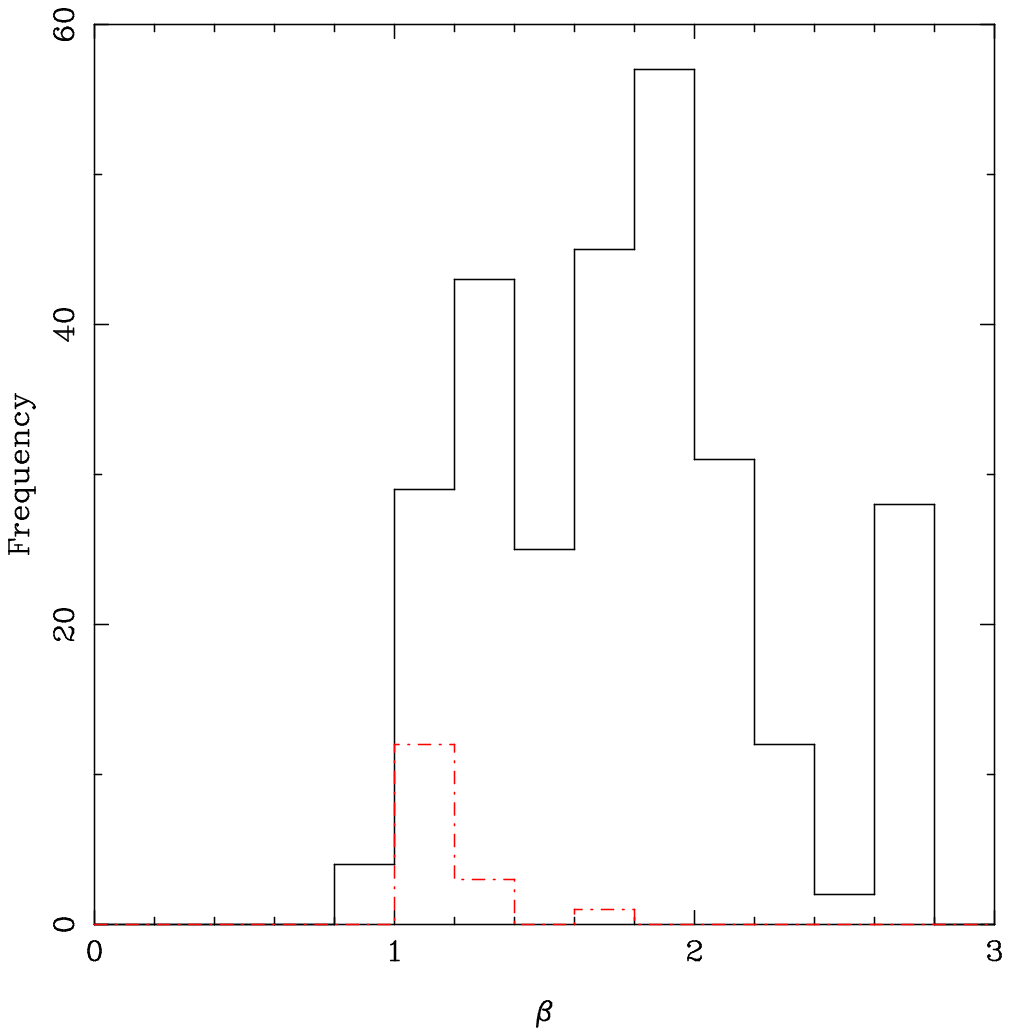}}}
\mbox{\subfigure{\includegraphics[width=0.23\textwidth,angle=-90]{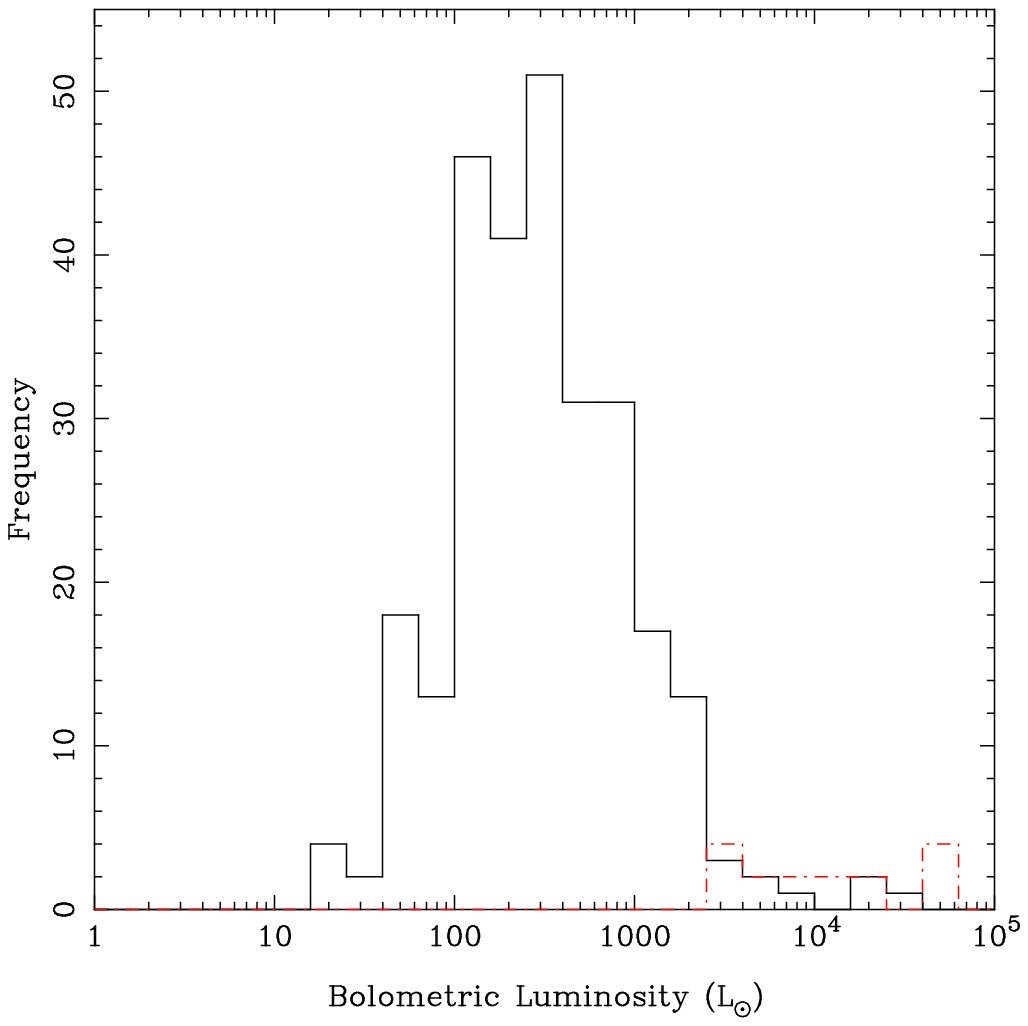}}
\subfigure{\includegraphics[width=0.23\textwidth,angle=-90]{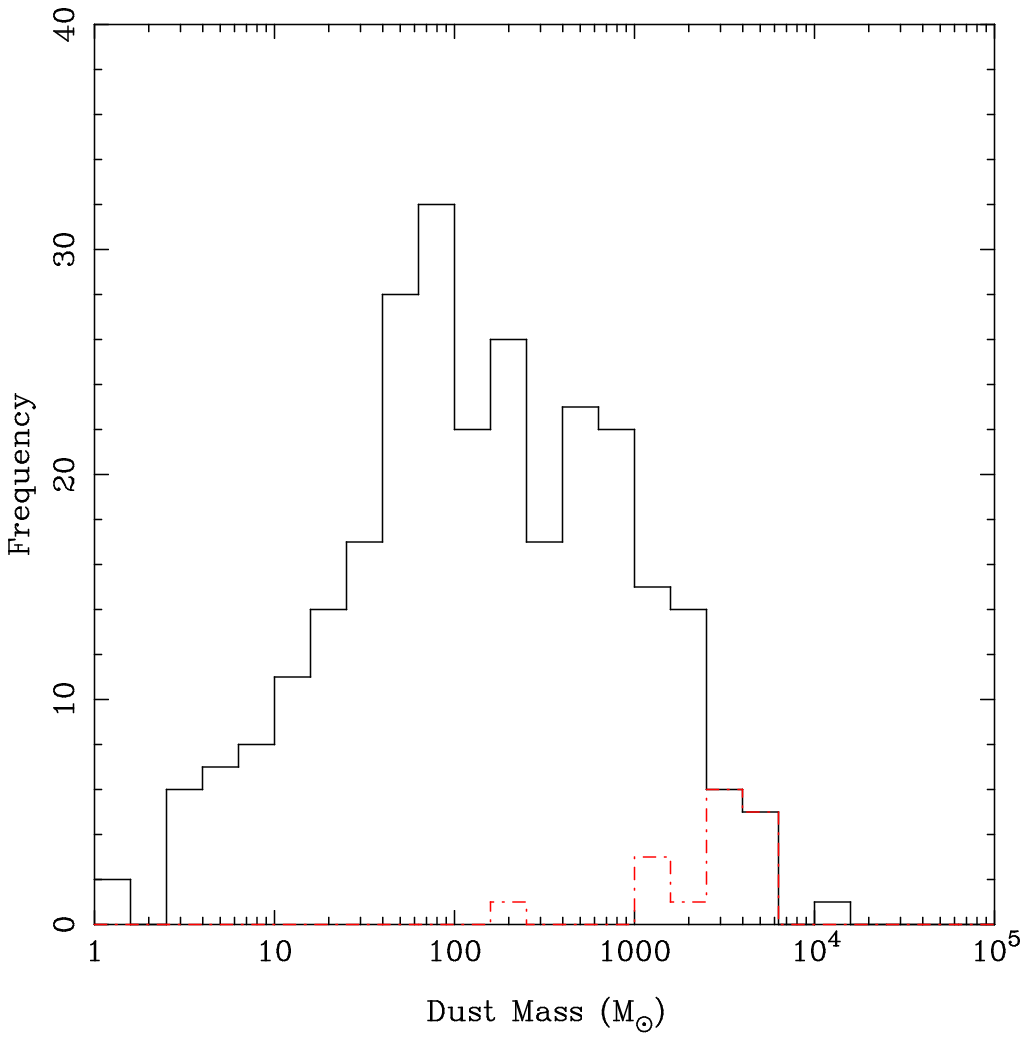}}}
\caption[dum]{Distribution of parameters for all sources with a reliable SED fit within G305 (solid black line), compared to the 16 identified candidate embedded massive star-forming regions (red dot-dashed line). \textit{Top left}: dust temperature. \textit{Top right}: dust emissivity index. \textit{Bottom left}: bolometric luminosity. \textit{Bottom right}: dust mass.}
\label{Figure:SED Parameter Histograms}
\end{figure}

In Fig. \ref{Figure:SED Parameter Histograms} we present the distribution of the modified blackbody parameters of the dust temperature, dust emissivity index, and the dust temperature, along with the bolometric luminosities obtained from \cite{Robitaille:2006mo} model SED fitting. We find that dust temperatures lie within the range of 10 to 42\,K, with a median value of $\approx$\,21\,K. For the dust emissivity index, a range between 0.8 to 2.8 is found, with the median of $\approx$\,1.8. The bolometric luminosity is sampled between 10 to 10$^{4}$\,\lsol, with a median of  $\approx$\,300\,\lsol, while the dust mass lies between 1 to 10$^{4}$\,\msol, with a median of  $\approx$\,540\,\msol.

\subsection{A Far-IR Selection Criterion For Embedded Massive Star Formation}

\begin{figure*}
\begin{center}
\includegraphics[width=0.65\textwidth,angle=-90]{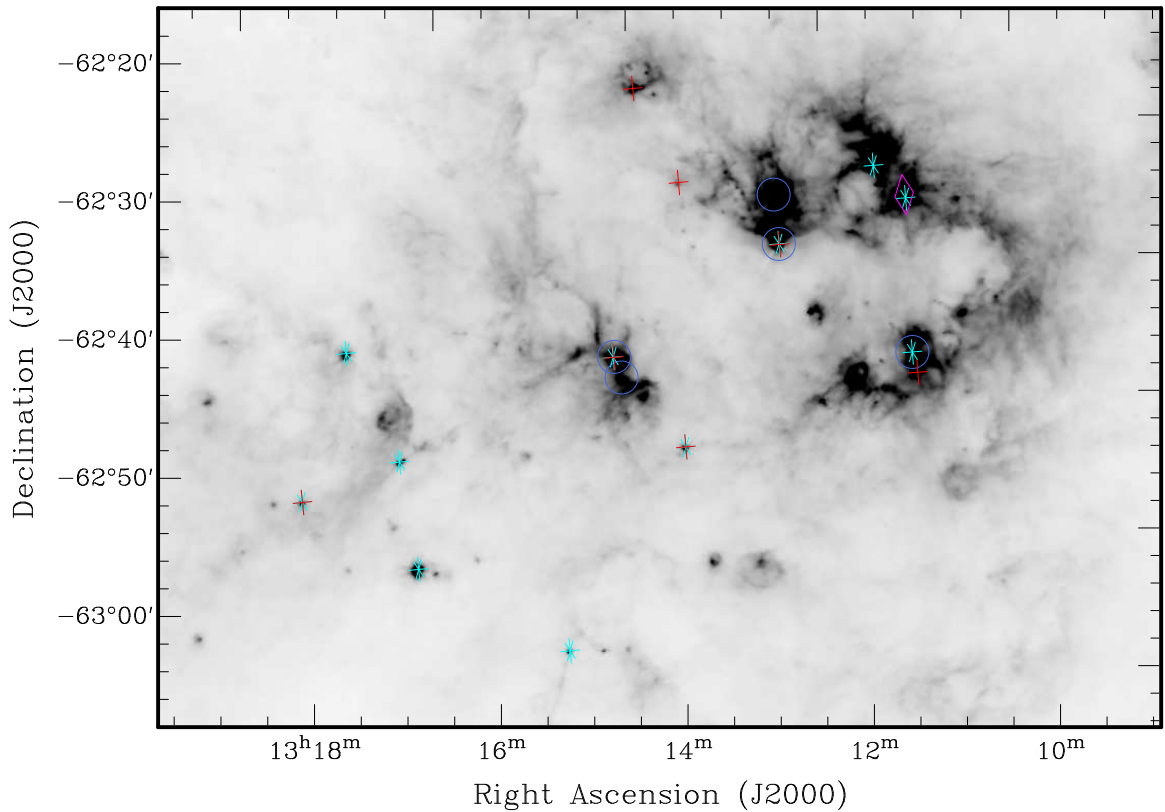} 
\caption{Location of candidate embedded massive star-forming regions and their relative association overlaid onto a 160\,$\mu$m greyscale Hi-GAL image; ATCA radio data (blue circles), MMB methanol masers (cyan asterisks), RMS sources (red crosses), H$_{2}$O masers (purple diamonds). Numerous regions are found to be associated with multiple star formation tracers.}
\vspace{-10pt}
\label{Figure:G305 Embedded Massive Star-Forming Associations}
\end{center}
\end{figure*}

Previous studies that have attempted to distinguish a sample of MYSOs have relied on colour selection criteria based on IRAS data \citep{Palla:1991kx,Molinari:1996uq,Walsh:1997sw,Sridharan:2002fj}, however these are restricted by selection effects, limitations in sensitivity and resolution, and also their incomplete Galactic coverage. The majority of these studies are biased in that they rely on bright IRAS sources, that with a resolution of 5\arcmin\,\,at 100\,\micron\,, are often found removed from the dense, confused regions along the Galactic plane where the majority of star formation is expected to be found. The result of these studies is a restricted sample that, due to selection issues, may not entirely represent the general MYSO population, making any extension to study massive star-formation problematic \citep{Urquhart:2008xr}.

It has been suggested that both the SED morphology, and the bolometric luminosity can prove effective in determining a sample of embedded massive star-forming objects \citep{Molinari:2008qy}. With the bulk of the YSO emission being in the far-IR portion of the SED, due to cold dust \citep{Molinari:2008tj}, the \textit{Herschel} Hi-GAL observations are ideal in accurately constraining the luminosity.

From the candidate associations identified, those sources found to have either a radio, MMB, H$_{2}$O maser, or RMS counterpart are known sites of massive star formation; tending to be luminous sources (i.e. $>$\,10$^{3}$\,L$_{\odot}$). Using this sub-sample we are able to identify a sample of known embedded massive star-forming regions from the physical properties derived from, and morphology of, their SEDs (shown to peak at 100\,\micron\,, as shown in Fig. \ref{Figure:Robitaille SED Fits}). We refer to this population of Hi-GAL sources that are associated with radio, MMB, H$_{2}$O maser or RMS counterparts as the embedded population. Added to this, we employ a selection cut in the bolometric luminosity of 10$^{3}$\,\lsol\,\,, that corresponds to the minimum spectral type that we define as a massive star (i.e. M\,$>$\,8\,M$_{\odot}$). In total we find some 16 candidate embedded massive star-forming regions, that match these selection criteria; their respective properties and relative associations are shown in Table \ref{tab:Embedded Massive Star-Forming Regions Parameters}. We also note the location of these associations, shown in Fig. \ref{Figure:G305 Embedded Massive Star-Forming Associations}, with the majority being located in the hot dust locations along the periphery of the central cavity, suggesting an interaction between the central sources of Danks 1 \& 2, and the surrounding material.

\begin{figure}
\begin{center}
\includegraphics[width=0.4\textwidth,angle=-90]{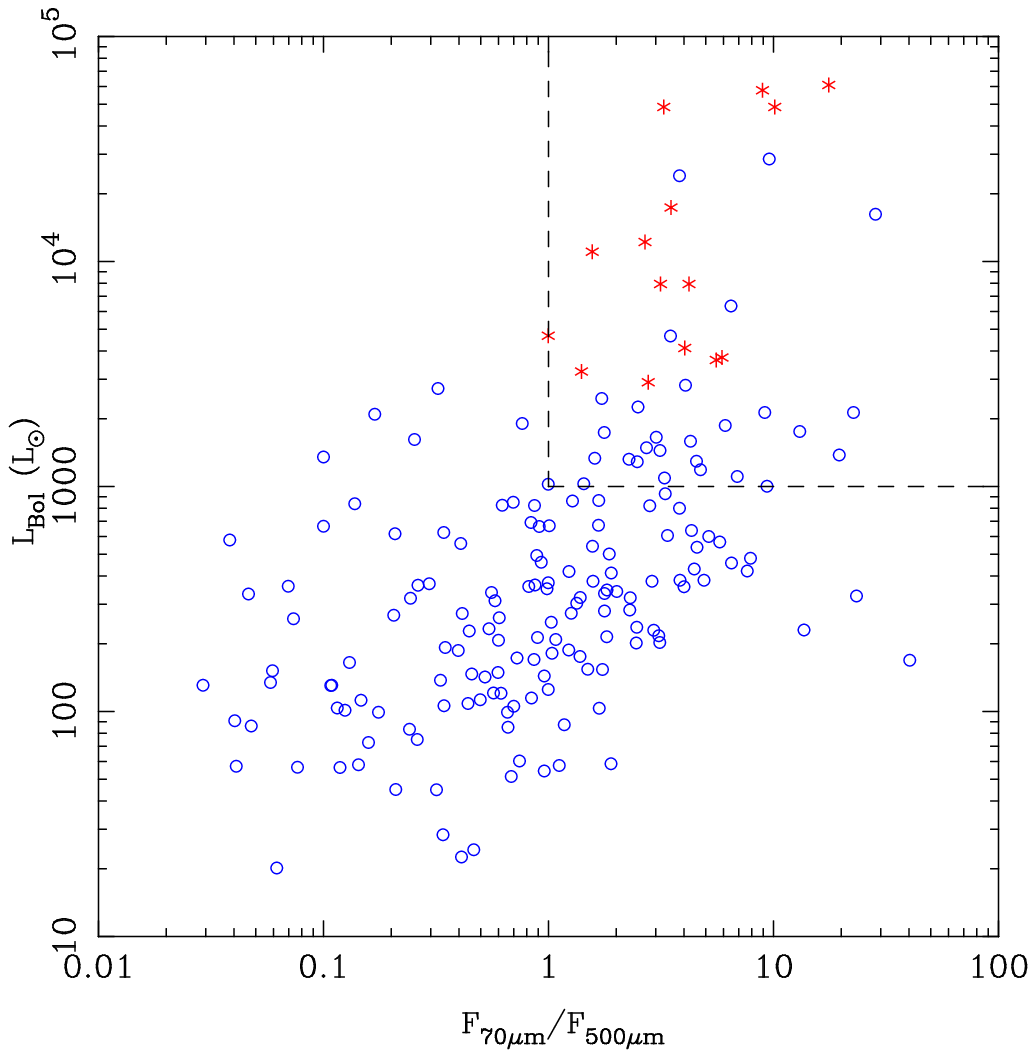} 
\includegraphics[width=0.4\textwidth,angle=-90]{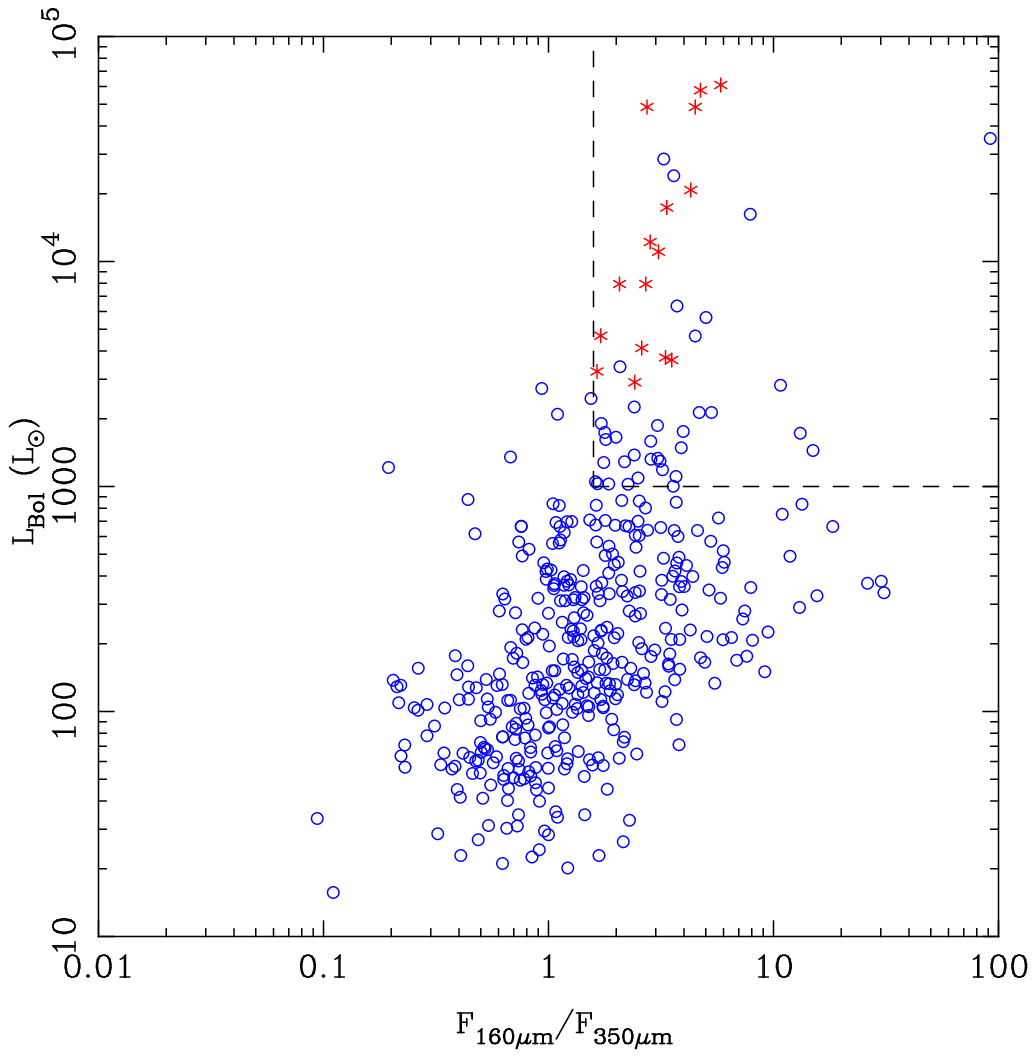} 
\caption{Colour - ­colour plots of all Hi-­GAL sources found in the G305 field (blue circles), and known sites of massive star formation (red asterisks). Dashed lines indicates the boundary of the region used for distinguishing sites of embedded massive star formation from other sources, at a luminosity of $>$\,10$^{3}$\,L$_{\odot}$.}
\label{Figure:FIR Colour Cut}
\end{center}
\end{figure}

Based on this sample of 16 embedded massive star-forming regions, we are able to derive a two-colour selection criterion to identify the overall embedded population of the G305 complex. We find that the 70\,-\,500\,\micron\,\,and 160\,-\,350\,\micron\,\,colours are most sensitive to the embedded population, shown as asterisks in Fig. \ref{Figure:FIR Colour Cut}. As can be seen from Fig. \ref{Figure:FIR Colour Cut}, the embedded population are tightly confined to one area of the colour plot, and can be distinguished from the remaining G305 population, shown as circles. From this we suggest a far-IR colour selection criterion for embedded massive star-forming regions of log\,(F$_{70}$/F$_{500}$)\,$\geq\,1$ and log\,(F$_{160}$/F$_{350}$)\,$\geq\,1.6$, yielding a further 31 embedded massive star candidates with no associated emission and luminosities $>$\,10$^{3}$\,L$_{\odot}$, as shown in Fig. \ref{Figure:FIR Colour Cut}. From these 31 candidates we find the faintest source to have a peak 70\,\micron\ flux of 1.02 Jy/pixel, and compared to the 90\% recovery rate of sources discussed in Sect. 2.1, we do not expect any further more deeply-embedded massive star-forming regions to be found within G305.

Currently, the nature of these 31 candidate embedded massive star-forming regions is unclear; from Fig. \ref{Figure:FIR Colour Cut} these candidates are predominately found with bolometric luminosities of $\approx$\,10$^{3}$\,\lsol\, however at least 3 sources are found a luminosities $\geq$\,10$^{4}$\,\lsol\ with no corresponding tracer of massive star formation. A lack of association to ATCA radio sources could be accounted for by localised noise found towards bright large-scale emission present with G305, confusing possible associations to real compact emission present (see \cite{Hindson:2012lr} Sect. 2.2 for a detailed discussion of the ATCA data reduction process). Another possibility could be the strong variability of both methanol \citep{Green:2012lr} and water masers \citep{Breen:2010eq}, to such an extent that they display no common features at the present epoch. Aside from the possible reasons for lack of associated tracers, this sample of candidates may also suggest an earlier, very young embedded phase present within G305. The possible nature of these additional candidates is particularly interesting, and warrants further investigation at a later date.

\begin{table}
\scriptsize
\addtolength{\tabcolsep}{-4pt}
\caption{Derived physical properties for all identified embedded massive star-forming regions, from both modified blackbody fits and \cite{Robitaille:2007nv} SED fitting technique, along with found associations.}
\begin{centering}
\begin{tabular}{lcccccc}
\hline
Hi-GAL & $\beta$ & T & $\tau_{500\mu m}$ & M$_{Dust}$ & L$_{Bol}$ & Association(s)\\
Source Index & & (K) & (10$^{-3}$) & (\msol) & (10$^{3}$\,\lsol) & \\
\hline
938   & 1.0 & 35 & 20.1 & 4300 & 48.6 &  MMB, H$_{2}$O maser\\
945   & 1.1 & 29 & 15.4 & 3900 & 7.95 & RMS \\
972   & 1.2 & 41 & 11.4 & 1200 & 20.8 & ATCA, MMB \\
1184 & 1.1 & 25 & 27.1 & 4700 & 11.1 & MMB \\
1800 & 1.4 & 35 & 29.4 & 3500 & 60.8 & ATCA, MMB, RMS \\
1804 & 1.0 & 32 & 24.8 & 5200 & 17.4 & ATCA \\
2114 & 1.7 & 24 & 1.29 & 200   & 2.91 & RMS \\
2153 & 1.2 & 29 & 9.14 & 3700 & 4.13 & MMB, RMS, MIPS \\
2212 & 1.1 & 23 & 14.3 & 4400 & 4.68 & RMS, MIPS \\
2363 & 1.1 & 27 & 14.2 & 3200 & 12.2 & ATCA, RMS \\
2383 & 1.0 & 36 & 20.0 & 1700 & 57.7 & ATCA, MMB, RMS \\
2627 & 1.4 & 28 & 4.18 & 1500 & 3.65 & MMB \\
2902 & 1.0 & 37 & 42.2 & 4200 & 48.6 & MMB\\
2923 & 1.1 & 24 & 16.1 & 3700 & 3.24 & MMB \\
2994 & 1.0 & 29 & 23.1 & 3900 & 7.95 & MMB, MIPS \\
3032 & 1.4 & 28 & 4.57 & 1500 & 3.75 & MMB, RMS\\
\hline
\end{tabular}
\label{tab:Embedded Massive Star-Forming Regions Parameters}
\end{centering}
\end{table}

\subsection{The Present-day Star Formation Rate Of The G305 Complex}

With a sample of embedded massive star-forming objects, one opportunity that is available is to study the recent star-forming history of the complex. Determining the star formation rate (SFR) at a local level is crucial in determining the global Galactic SFR, helping to unveil any mechanisms that may lead to global scaling laws \citep{Molinari:2010gl}. The SFR, along with the IMF, express the population of massive stars within the Galaxy, and determine what the impact on the local environment is, such as the composition of the ISM, the rate of feedback from massive stars, and the rate of conversion of gas into stars \citep{Calzetti:2010aa}.

Given our position within the Galactic disk, direct SFR indicators using optical/UV tracers will fail to reproduce an accurate SFR due to high extinction rates of the dusty ISM. However, far-IR observations, unaffected by extinction, provide us with the ability to resolve the YSO population associated with HII regions, allowing constraints on the IMF and stellar ages, yielding a detailed star formation rate of Galactic HII regions \citep{Chomiuk:2011fk}.

A SFR derived from a resolved, YSO counting approach, or from that inferred from the total infrared luminosity has to assume a `steady-state' approximation to reliably trace the star formation activity of the region. The assumption in these calculations is that both the rate of massive star formation, and the rate that massive stars evolve off the main sequence, is in approximate equilibrium \citep{Krumholz:2012lr}. For this to be true, the requirement is that the age of the region be longer than that of the UV emitting population used to trace the SFR \citep{Kennicutt:1998fk}. We show below that this is the case for a realistic star-forming timescale.

\subsubsection{The Embedded Massive Star Formation Rate}

With our sample of 16 identified embedded massive star-forming regions, and the further 31 embedded massive star candidates found, we are able to comment on the present star formation history of the G305 complex. If we make the assumption that for each embedded massive star-forming region identified, the most massive star present produces the majority of the bolometric luminosity and is also accompanied by a cluster of lower mass stars, we are able to scale the IMF accordingly; for our calculations, we adopt a simple IMF proposed by \cite{Salpeter:1955xv}. By comparing the calculated bolometric luminosity for each candidate region found, to the luminosities calculated for a sample of MYSOs from \cite{Mottram:2011fs}, we are then able to estimate the most massive star for each region. 

In order to extrapolate the IMF from observed mass to lower mass, we are obliged to select both a lower and upper mass limit that all observed YSOs fall within the selected range. For our purposes we adopt a lower mass of 0.1\,\msol\,\,and an upper limit of 50\,\msol, as used when calculating the Galactic SFR \citep{Robitaille:2010eq}.

By adopting a Salpeter IMF, which best samples the high-mass tail of the IMF \citep{Zinnecker:2007mk}, and assuming a constant SFR therefore, we arrive at $\approx$\,10$^{4}$ YSOs present, which corresponds to a total mass in stars of $\approx\,8\,\times$\,10$^{3}$ M$_{\odot}$. Since we consider the total mass in stars, we select a typical timescale for that mass to assemble, which would be the time to reach the pre main-sequence of 0.5\,Myr \citep{Offner:2011ph}, from this we attain a SFR for G305 of $\approx$\,0.01\,-\,0.02 M$_{\odot}$\,\,yr$^{-1}$. In this scenario, the `steady-state' approximation should hold since the timescales of both MYSOs, $\approx$\,10$^{4}$\,yr \citep{Mottram:2011fs}, and UC HII regions, $\approx$\,10$^{5}$\,yr \citep{Comeron:1996fr}, are some 5\,-\,50 times shorter than our assumed timescale for the star formation within G305 of $\approx$\,0.5\,Myr.

A similar approach has been taken in \cite{Hindson:2012lr}, using the identified UC HII population (some five identified in total) of G305 to derive a SFR, over the last 0.5\,Myr, of 0.002\,-\,0.004 M$_{\odot}$\,\,yr$^{-1}$. This rate is considered a lower limit due to the incompletenesses in the ATCA radio data, with a \textit{uv} cut implemented to emphasise the compact, small scale radio emission associated with UC HII regions. For comparison, \cite{Davies:2011gl}, have determined the SFR for the G305 complex over the last 5\,Myr from the calculated age and mass of the two central open clusters Danks 1 \& 2. The SFR was found to be $\approx$\,0.002\,-\,0.005 M$_{\odot}$\,\,yr$^{-1}$, which is comparable to our derived SFR from counting the embedded YSO population. However, taking our derived SFR of $\approx$\,0.01\,-\,0.02 M$_{\odot}$\,\,yr$^{-1}$, it is clear that the star formation activity of G305 has not remained constant over the last 5\,Myr. If this were to be the case, then we would expect to observe some 75\,000 M$_{\odot}$ of stars to be observed within the complex, which is entirely not the case. The `collect and collapse' model of star formation, proposed by \cite{Elmegreen:1977bs}, requires a period of time after the formation of the central ionising source(s) for material within the surrounding molecular clouds to be swept-up by the expansion of the HII region. If this expansion continues for a sufficient time, the swept-up shell of material becomes self-gravitating and is then expected to fragment, entering a phase of collapse possibly leading to the formation of new stars. Following the approach taken by \cite{Dale:2007nu}, who have calculated this `fragmentation time-scale' for a uniform molecular cloud of pure molecular hydrogen, we take the ionising photon flux of the G305 complex from \cite{Clark:2004vd} to estimate a similar timescale. We find that after the formation of Danks 1 \& 2, there would be a delay of $\approx$\,2.4\,Myr until the next generation of star formation occurred within the complex. This would support the scenario that the star formation within G305 was not continuous, but more likely characterised by punctuated star formation over the lifetime of the complex.

We find also that our derived SFR for G305 is comparable to other well known massive star-forming complexes in the Galaxy, namely the Carina complex \citep{Povich:2011oj}, and M17 \citep{Chomiuk:2011fk}. We stress that the derived SFR value is based on a small sample of high mass stars, and has been extrapolated over a large range of stellar masses; when considering the lower mass stars present, their lifetimes may well be 1-2 orders of magnitude longer. 

For completeness, the Galactic SFR is found to be $\approx$\,2 M$_{\odot}$\,\,yr$^{-1}$ \citep{Chomiuk:2011fk, Davies:2011af}, suggesting that a few tens to hundreds of G305 complexes are analogous to the entire star formation rate of the Milky Way. Using results from the Wilkinson Microwave Anisotropy Probe (WMAP), \cite{Murray:2010fk} identify some 31 Galactic HII regions with an ionising flux greater than that of G305, of which some 18  WMAP sources constitute over half the total Galactic  ionising flux. Just as the IMF is dominated by the more massive stars present, the Galactic SFR is probably dominated by the few rigorous star-forming regions present.

\subsection{Alternative Star Formation Rate Indicators Within G305}

Clearly the value for the SFR derived from the population of embedded massive star-forming regions identified will be an upper limit, since we have assumed a power-law slope (i.e. Salpeter IMF), and by extrapolating the IMF over a small sample of massive stars have overestimated the total mass in stars. To this must also be added issues of completeness, and assuming a timescale of $\approx\,10^{5}$\,yrs  will be unrepresentative of the intermediate to low-mass YSOs present. However the value should prove a good assumption for the upper limit, that can then be compared to other SFR tracers that are independent of the resolved massive stellar population within G305.

What is key here is that we are able to resolve the YSO population within Galactic HII regions, such as G305, and use both the IMF and stellar timescales to derive a SFR. This can then be contrasted and calibrated to extragalactic emission tracers, such as the total IR luminosity, to determine whether Galactic SFRs are consistent with extragalactic SFR indicators. We next consider the SFR using tracers that are independent of the identified YSO population. Table \ref{tab:SFR Indicators} lists the calculated SFRs for G305 using numerous tracers, with reference to each approach.

\subsubsection{The Relation Between Star Formation Rate and Molecular Cloud Mass}

Recent work by \cite{Heiderman:2010io} and \cite{Lada:2010tb} on the star formation activity of molecular clouds within 0.5\,kpc of the Sun, suggest that the star formation rate scales linearly with the molecular cloud mass. \cite{Lada:1992fk} showed that active star formation is to be found primarily in high volume density regions of molecular clouds, with star formation favouring very massive, dense cores. The expectation for the star formation activity is thus that there is a tight correlation with the amount of high extinction material present in molecular clouds.

\cite{Lada:2010tb} illustrate for a sample of local molecular clouds, that by comparison of the cumulative mass to YSO content as a function of extinction (see \cite{Lada:2010tb} Fig. 3), a marked minimum dispersion of the cumulative mass is found at A$_{V}$\,=\,7.3\,$\pm$\,1.8 mag. The proposition is that above this minimum, the cloud mass is directly related to the star formation activity and hence the SFR within the clouds. It is shown that this high extinction value corresponds to an equally high volume density of n(H$_{2}$)\,$\approx$\,10$^{4}$\,cm$^{-3}$ \citep{Lada:2010tb}. There is also evidence this linear relation holds for extragalactic molecular clouds. A tight correlation between the total IR luminosity and the luminosity of the HCN molecule, which itself requires high densities ($\textgreater$\,10$^{4}$\,cm$^{-3}$) to be excited to a detectable level, has been found for both Galactic cores \citep{Wu:2005fk}, and for a sample of normal spirals and starburst galaxies \citep{Gao:2004fk}. These studies suggest that the linear relation holds for dense interstellar gas both on a Galactic and extragalactic scale, underlying a physical relation that links star formation and galaxy evolution.

From this linear correlation, a SFR is derived of the form:

\begin{equation}
SFR=4.6\,\pm\,2.6\,\times\,10^{-8}\,M_{0.8}\,\,\rm{[M_{\odot}\,\,yr^{-1}]}
\label{eq:Dense gas SFR equation}
\end{equation}
where M$_{0.8}$ corresponds to the cloud mass, in solar masses, above an extinction threshold of A$_{K}$\,$\approx$\,0.8 mag, which is derived from the visual extinction of A$_{V}$\,=\,7.3\,$\pm$\,1.8 mag mentioned earlier.

For each IR source identified by \textit{Herschel} across the G305 complex we determine the physical radius at 250\,\micron, as this wavelength offers the optimum combination of signal-to-noise, and angular resolution. To estimate the physical diameter for each source, we firstly deconvolve the source size from the Gaussian beam \citep{Thompson:2004qy}:

\begin{equation}
\Theta^{2}_{\rm{Source}}=\Theta^{2}_{\rm{Obs}} - \Theta^{2}_{\rm{Beam}}
\label{eq:Convolution Equation}
\end{equation}
where $\Theta^{2}_{Obs}$ is the source size estimated from the FWHM of the Gaussian fitting, and $\Theta^{2}_{Beam}$ is the \textit{Herschel} beam size at 250\,\micron. If we then place each source at a distance of 4\,kpc, and assume spherical geometry and a uniform density, we can determine those sources found to be above the critical density threshold. From this we are able to determine the mass of the dense gas within G305, then using the \cite{Lada:2010tb} assumption that dense gas is associated with the star formation activity, we find a dense gas mass of $\approx$\,3\,$\times$\,10$^{5}$\,M$_{\odot}$; for comparison, the total molecular mass traced by NH$_{3}$ is found to be $\approx$\,6\,$\times$\,10$^{5}$\,M$_{\odot}$ \citep{Hindson:2010jt}. We note that this mass is an approximation, since it is unlikely that the density of each source is uniform. 

By combining this mass with Equation (\ref{eq:Dense gas SFR equation}), we obtain a dense gas derived SFR of 0.006\,-\,0.02 M$_{\odot}$\,\,yr$^{-1}$. The result is found to be in good agreement with the embedded massive star formation rate derived earlier, and goes some way toward confirming the \cite{Lada:2010tb} assumption by extension to the star formation activity of embedded massive star-forming regions. 

\subsubsection{The 70\,\micron\ Emission Star Formation Rate}

In contrast to deriving a value of the SFR from all identified embedded massive star-forming regions, we can also approach measuring the SFR by considering the total infrared flux (TIR) of the giant G305 HII region. On an extra-galactic perspective current SFRs are calculated using a tracer of UV photon emission from YSOs, and also spectral synthesis models \citep{Kennicutt:1998fk}. In this case, observations of HII regions prove ideal measures of current star formation. However, some fraction of the total UV emission will be obscured by the presence of dust, thus bolometric IR observations of dust (i.e. TIR) will provide an excellent means to recover the extinguished UV photon emission, with the dust absorption highly peaked in the UV and re-emission being in a broad spectral range of mid-to-far-IR \citep{Kennicutt:1998fk}. The conclusion from this is that the TIR will provide the best indicator of SFR obscured by the presence of dust.

Observationally, an advantage would be the use of a single-band star formation indicator, such as UV, H$\alpha$, 8\,\micron, 24\,\micron\,\,etc., however each of these has its own complications. In the case of UV and optical lines, corrections need to be taken into consideration due to large extinction. Whereas 8 and 24\,\micron\,\,emission strongly depends on the local environment, since the abundances of small dust grains that contribute to their emission depend greatly on metallicity and the presence of ionising radiation. 

\cite{Calzetti:2007xx} and \cite{Dale:2005sz} note that 8\,\micron\,\,emission makes for an inaccurate SFR indicator since there is a large degree of variability of emission in galaxies with respects to SED shape and metallicity.  This strong variability at 8\,\micron\,\,is emphasised in Fig. \ref{Figure:Robitaille SED Fits}, where variations of an order of magnitude exist between the best fit and good fit SED models. This variation can be accounted for by the disk inclination to the line of sight for the centrally embedded object, where the observed flux from a pole-on view can be 2\,-\,4 times greater than a more edge-on viewing angle \citep{Whitney:2003zr}.

\cite{Calzetti:2005oh} find that the SFR calculated from 24\,\micron\,\,emission itself varies strongly from galaxy to galaxy. On a local scale, the ratio of the 24\,\micron\,\,luminosity to SFR is found to be a reasonably accurate tracer, however when applied to other systems, such as starbursts and ultraluminous infrared galaxies (ULIRGs), the ratio is systematically higher. This variation in the 24\,\micron\,\,is found to be a factor of a few with respects to the observed SEDs, and may be grounded in the strong dependence on local galactic conditions; with ionising stars heating dust to different averaged temperatures, the 24\,\micron\,\,emission will be most sensitive to this. However, \cite{Dale:2005sz} find that 70\,\micron\,\,emission may be an accurate monochromatic star formation indicator, since the 70-to-160\,\micron\,\,ratio is found to correlate well with local SFRs.

Recent work by \cite{Lawton:2010rj} have determined an accurate monochromatic IR band that best approximates the obscured SFR in the Large Magellanic (LMC) and Small Magellanic Clouds (SMC), through IR aperture photometry of 16 LMC and 16 SMC HII regions, using \textit{Spitzer} IRAC (3.6, 4.5, 8\,\micron) and MIPS (24, 70, 160\,\micron) bands. It is found from the IR SEDs of each HII region, that the majority peak at around 70\,\micron\,\,at all radii (10\,-\,400\,pc) from the centrally ionising sources, and that the 70\,\micron\,\,emission most closely traces the size of each HII region as found using the TIR. The conclusion from this is that the 70\,\micron\,\,emission is the most likely suitable IR band to utilise as a monochromatic SFR indicator.

It has been argued by \cite{Kennicutt:1998fk} that the TIR is the best obscured SFR indicator available for starburst galaxies. However, dust obscured star formation in HII regions are found to behave similarly, in that their environments are both very dusty and are sites of recent star formation. The \cite{Kennicutt:1998fk} obscured SFR equation is of the form:

\begin{equation}
SFR=4.5\,\times\,10^{-44}\,L_{\rm{TIR}}\,\,\rm{[M_{\odot}\,\,yr^{-1}]}
\label{eq:Kennicutt SFR Equation}
\end{equation}
where L$_{TIR}$ is the TIR luminosity in erg\,\,s$^{-1}$, and the value 4.5\,$\times$\,10$^{-44}$ is a constant derived from synthesis models, with assumptions on the IMF and star formation timescales \citep{Kennicutt:1998fk}.

The TIR luminosity in Equation (\ref{eq:Kennicutt SFR Equation}) can be substituted with the averaged 70\micron\,\, luminosity, normalised by the TIR, while also applying an IR band specific constant. The monochromatic obscured SFR equation of \cite{Lawton:2010rj} is found to be:

\begin{equation}
SFR=9.7(0.7)\,\times\,10^{-44}\,L(\lambda)\,\,\rm{[M_{\odot}\,\,yr^{-1}]}
\label{eq:Monochromatic obscured SFR}
\end{equation}
where L($\lambda)$ is the observed luminosity in erg\,\,s$^{-1}$.

By employing aperture photometry of the whole G305 region, we obtain the cumulative 70\,\micron\,\, flux, f$_{\nu}(\lambda)$, and from this are able to calculate the monochromatic luminosity at 70\,\micron\,\,\citep{Calzetti:2010zh}:

\begin{equation}
L(\lambda)=4\,\pi\,d^{2}\,\left(\frac{c}{\lambda}\right)\,f_{\nu}(\lambda)\,\,\rm{[erg\,\,s^{-1}]}
\label{eq:Monochromatic luminosity}
\end{equation}
where d is the distance to the G305 complex, in m. 

Using the value found for the observed luminosity at 70\,\micron\,, with Equation (\ref{eq:Monochromatic obscured SFR}), we obtain an obscured SFR for G305 of 0.002\,-\,0.005 M$_{\odot}$\,\,yr$^{-1}$. A similar approach has been suggested by \cite{Li:2010hr}, who also determine a monochromatic SFR indicator at 70\,\micron\,, yet calibrate their SFR tracer not with the TIR luminosity but rather with the combined H$\alpha$, and 24\,\micron\,\,luminosity. For completeness, using the \cite{Li:2010hr} tracer, we derive a SFR of 0.004\,-\,0.008 M$_{\odot}$\,\,yr$^{-1}$, which is in approximate agreement with that using the \cite{Lawton:2010rj} approach.

We can directly compare these results to that derived from the total Lyman continuum photon rate of G305, where we find an SFR of 0.002\,-\,0.004\,M$_{\odot}$\,\,yr$^{-1}$ (Hindson et al. in prep). We note that both these two independent tracers are in excellent agreement, however are found to a factor of $\geq$\,2 lower than that derived from the embedded massive star-forming population. A similar result is found by \cite{Chomiuk:2011fk}, who find that the SFR for M17 estimated from both the Lyman continuum and 24\,\micron\,\,emission is underestimated by a factor of $\geq$\,2 in comparison to the SFR derived from YSO counting. 

This discrepancy may simply be that there are elementary differences in the measurements between Galactic and extragalactic observations. \cite{Lawton:2010rj} note that Equation (\ref{eq:Monochromatic obscured SFR}) extends to HII regions measured at projected distances of 52\,kpc and 61\,kpc for the LMC and SMC respectively. With G305 some $\sim$\,4\,kpc distant, the relation between the SFR and luminosity at 70\,\micron\,\,may indeed break down, with the effects of individual protostars becoming more important, due to the larger spatial resolution. It may also be the case that the `steady-state' assumption breaks down in this comparison. Though it is normally true that the lifetime of the region observed is longer than for the individual objects for extragalactic realms, this may not hold for observations of Galactic regions that tend to be smaller, and with shorter dynamical timescales.

\subsubsection{A Galactic - Extragalactic Comparison}

By comparing the derived SFR from numerous tracers, shown in Table \ref{tab:SFR Indicators}, what is immediately apparent is the disparity between the rates derived from the resolved stellar population and those from extragalactic tracers; there is lack of consistency between the two, with extragalactic tracers tending to underestimate the SFR derived from resolved Galactic SFRs. This circumstance between the two SFR regimes has been noted by several authors \citep{Heiderman:2010io,Lada:2010tb,Chomiuk:2011fk}, where there appears a distinct underestimation for Galactic H II regions. \cite{Heiderman:2010io} make the suggestion that this difference may be accounted for by the inclusion of diffuse gas, in the standard Kennicutt-Schmidt relation, that is below the critical density threshold for star formation, as suggested by \cite{Lada:2010tb}. Since extinction maps are not readily available to determine the surface density of gas in extragalactic studies, CO maps are often employed instead. \cite{Heiderman:2010io} find that using CO as a gas tracer, for a sample of local molecular clouds, leads to an underestimate in mass of  $\gtrsim$\,30\,\% compared to that obtained using extinction maps. The result of this would essentially push down the estimated SFR from the \cite{Kennicutt:1998fk} relation, and may go some way in accounting for the dissimilarity between extragalactic regions and more local, Galactic ones.

However, recent work by \cite{Krumholz:2012lr} seems to suggest a unified star formation law, with objects ranging from both low mass Solar neighbourhood clouds through to sub-mm galaxies in agreement with one distinct star formation law. What is advocated in this law is that the SFR, within a variety of scales, is simply $\approx$\,1\,$\%$ of the molecular gas mass per local free-fall time. This volumetric approach suggests a local, universal star formation law that is applicable to Galactic and extragalactic observations (see \cite{Krumholz:2012lr} Fig. 3), bridging the gap between the apparent disparity in the two regimes. This law is affected solely by local variations, such as the gas condition, with more global Galactic/galaxy-scale properties, such as the orbital period, having no impact on the SFR in so far as they do not change the local properties of star-forming regions. 

Conversely, \cite{Lada:2011lr} conclude that a universal star formation law that is applicable from the Milky Way through to near-IR selected (BzK) galaxies, is simply directed by the amount of dense molecular gas that can accumulate within a star-forming region. In the majority of situations, only 10\,$\%$ of the total mass within a molecular cloud is at a sufficient density, n(H$_{2}$)\,$\geq$\,10$^{4}$\,cm$^{-3}$ \citep{Lada:2010tb}, to actively form stars. Clearly there is a disparity between the two proposed universal SFR laws; \cite{Krumholz:2012lr} favouring gas surface densities and local free-fall times as crucial, while \cite{Lada:2011lr} advocate gas surface densities and the fraction of dense gas as the pivotal factors. It therefore seems that more work is needed to describe the underlying nature of a universal star formation law, if such a law is to be found.

What is apparent, when measuring the Galactic SFR, is the need for an accurate means to compare the Milky Way to other galaxies, in order to allow us to extend the more detailed Galactic analysis to other systems and to test the discrepancy between the two regimes. Continued, multiwavelength analysis of Galactic HII regions, now including \textit{Herschel} Hi-GAL, will in part aid with this. Through extended study across a wide range of star-forming regions, an accurate determination of the IMF, and with that the SFR, can be achieved, which scaled up from a more local level to a global, Galactic level, will allow for the consideration of how these crucial properties vary as a function of environment across the Milky Way (see \cite{Veneziani:2012fk} for a detailed study of the Herschel Science Demonstration Phase fields). This should help in a better understanding of how the SFR can accurately be measured on both Galactic, and extragalactic scales, and lead to a more unified calibration.

\begin{table}
\scriptsize
\addtolength{\tabcolsep}{-5pt}
\caption{Calculated SFR for G305 using multiple SFR tracers.}
\begin{centering}
\begin{tabular}{lcc}
\hline
SFR Tracer & SFR & Reference \\ & (M$_{\odot}$\,\,yr$^{-1}$) & \\
\hline
Embedded Massive Stars & 0.01\,-\,0.02 & This Paper \\
Dense Gas & 0.006\,-\,0.02 & \cite{Lada:2010tb} \\
& & \\
UC HII Regions & $\geq$\,0.002\,-\,0.004 & \cite{Hindson:2012lr} \\
Danks 1 \& 2  & 0.002\,-\,0.005 & \cite{Davies:2011gl} \\
70\,\micron\,\, Emission & 0.002\,-\,0.005 & \cite{Lawton:2010rj} \\
 & 0.004\,-\,0.008 & \cite{Li:2010hr} \\
Lyman Continuum & 0.002\,-\,0.004 & Hindson et al. in prep.\\
\hline
\end{tabular}
\label{tab:SFR Indicators}
\end{centering}
\end{table}

\section{Summary}

We have studied the G305 star-forming complex, using \textit{Herschel} Hi-GAL far-IR data in search of an embedded massive star-forming population. In total, we identify some 16 embedded massive star-forming regions from their associations to radio, maser, and RMS counterparts across the region (Fig. \ref{Figure:G305 Embedded Massive Star-Forming Associations}). From this sample of known embedded massive stars we suggest a far-IR colour selection of log\,(F$_{70}$/F$_{500}$)\,$\geq\,1$ and log\,(F$_{160}$/F$_{350}$)\,$\geq\,1.6$ that can be utilised across similar regions within the \textit{Herschel} Hi-GAL survey, to identify embedded massive star-forming candidates across the Galactic plane.

With our sample of embedded massive stars, we derive the present-day SFR for the complex of $0.01\,-\,0.02\,\rm{M_{\odot}\,\,yr^{-1}}$, which is found to be in good agreement with other well known massive star-forming complexes such as the Carina complex \citep{Povich:2011oj}, and M17 \citep{Povich:2010zx}. In comparison to other well known extragalactic SFR tracers, based on the \cite{Kennicutt:1998fk} relation, there is a noted discrepancy between the two regimes, with extragalactic tracers tending to underemphasise the SFR. We note however, the use of the \cite{Lada:2010tb} relation for SFR that is in good agreement with the resolved SFR derived from our study, suggesting the key to the SFR is the total amount, and production of, dense gas within molecular clouds.

We find that the Milky Way SFR \citep{Chomiuk:2011fk, Davies:2011af} is comparable to tens to hundreds of G305 complexes, emphasising the fact that the Galactic SFR is most likely dominated by similar massive star-forming complexes. By combining \textit{Herschel} Hi-GAL data \citep{Molinari:2010gl} with current and future multi-wavelength large datasets such as GLIMPSE \citep{Benjamin:2003qy}, UKIDSS GPS \citep{Lucas:2008lr}, MIPSGAL \citep{Carey:2009qa}, VVV \citep{Minniti:2010fk}, CORNISH \citep{Purcell:2010fj}, and MALT90 \citep{Foster:2011lr} we will be able to study massive star formation across the Galactic plane, and how the SFR of numerous complexes vary as a function of environment. The result could conceivably give rise to global star-formation scaling laws that could then allow us to extend our detailed knowledge of our Galaxy to more distant extragalactic realms.

\section*{Acknowledgements}

A. Faimali would firstly like to thank the Science and Technology Facilities Council for a studentship.

The authors would firstly like to thank the entire \textit{Herschel} Hi-GAL team for their continuing work on the survey. We would also like to thank Sean Carey and Sachin Shenoy for producing, and making available the MIPSGAL point source catalogue for inclusion in our analysis. This research has made possible with the use of the NASA Astrophysics Data System Bibliographic Services. This paper made use of information from the Red MSX Source survey data base at www.ast.leeds.ac.uk/RMS, which was constructed with support from the Science and Technology Facilities Council of the UK. We made use of positional data on 6.7\,GHz methanol masers provided by the Methanol Multi-Beam (MMB) survey, also supported by the Science and Technology Facilities Council, and with a maser data base available at http://astromasers.org.
 
The \textit{Herschel} spacecraft was designed, built, tested, and launched under a contract to ESA managed by the \textit{Herschel}/\textit{Planck} Project team by an industrial consortium under the overall responsibility of the prime contractor Thales Alenia Space (Cannes), and including Astrium (Friedrichshafen) responsible for the payload module and for system testing at spacecraft level, Thales Alenia Space (Turin) responsible for the service module, and Astrium (Toulouse) responsible for the telescope, with in excess of a hundred subcontractors.
 
PACS has been developed by a consortium of institutes led by MPE (Germany) and including UVIE (Austria); KU Leuven, CSL, IMEC (Belgium); CEA, LAM (France); MPIA (Germany); INAF-IFSI/OAA/OAP/OAT, LENS, SISSA (Italy); IAC (Spain). This development has been supported by the funding agencies BMVIT (Austria), ESA-PRODEX (Belgium), CEA/CNES (France), DLR (Germany), ASI/INAF (Italy), and CICYT/MCYT (Spain).
 
SPIRE has been developed by a consortium of institutes led by Cardiff University (UK) and including Univ. Lethbridge (Canada); NAOC (China); CEA, LAM (France); IFSI, Univ. Padua (Italy); IAC (Spain); Stockholm Observatory (Sweden); Imperial College London, RAL, UCL-MSSL, UKATC, Univ. Sussex (UK); and Caltech, JPL, NHSC, Univ. Colorado (USA). This development has been supported by national funding agencies: CSA (Canada); NAOC (China); CEA, CNES, CNRS (France); ASI (Italy); MCINN (Spain); SNSB (Sweden); STFC (UK); and NASA (USA).

\bibliographystyle{mn2e.bst}
\bibliography{Paper_Reference.bib}{}
\bibstyle

\label{lastpage}
\end{document}